\pdfoutput=1 
\documentclass[10pt,twocolumn]{article}
\usepackage[a4paper,margin=2cm,columnsep=0.8cm]{geometry} 
\usepackage[T1]{fontenc}
\usepackage[utf8]{inputenc}
\usepackage{lmodern}
\usepackage{microtype}

\usepackage{amsmath,amssymb,amsthm}
\usepackage{graphicx}
\usepackage{booktabs}
\usepackage{xcolor}
\usepackage{braket} 

\usepackage{authblk}

\usepackage[numbers,sort&compress]{natbib}
\usepackage[colorlinks=true,linkcolor=blue!60!black,citecolor=blue!60!black,urlcolor=blue!60!black]{hyperref}
\usepackage{orcidlink} 

\theoremstyle{definition}

\makeatletter
\let\@savedthanks\@empty
\let\@origthanks\thanks
\renewcommand{\thanks}[1]{\@origthanks{#1}%
  \protected@xdef\@savedthanks{\@savedthanks\protect\footnotetext[\the\c@footnote]{#1}}}
\newcommand{\saythanks}{\begingroup\renewcommand{\thefootnote}{\fnsymbol{footnote}}%
  \@savedthanks\endgroup\global\let\@savedthanks\@empty}
\makeatother
\title{Spontaneous emission of atomic systems near metallic and semiconducting atomic chains}

\author[1,2]{Abhishek Ghosh\orcidlink{0000-0001-8350-1158}\thanks{Corresponding author: \texttt{ov@doktorant.umk.pl}}}
\author[3]{David Dams\orcidlink{0009-0001-8944-2643}}
\author[4,5]{Antton Babaze\orcidlink{0000-0002-9775-062X}}
\author[6,7]{Garnett Bryant\orcidlink{0000-0002-2232-0545}}
\author[3,8]{Carsten Rockstuhl\orcidlink{0000-0002-5868-0526}}
\author[4,9]{Andr\'{e}s Ayuela\orcidlink{0000-0001-9611-9556}}
\author[1,2]{Karolina S\l{}owik\orcidlink{0000-0003-1314-7004}\thanks{Corresponding author: \texttt{karolina@umk.pl}}}
\author[1,2]{Marta Pelc\orcidlink{0000-0002-0671-4990}}

\affil[1]{Institute of Physics, Nicolaus Copernicus University in Toruń, Grudziadzka 5/7, 87-100 Toruń, Poland}
\affil[2]{Institute of Advanced Studies, Nicolaus Copernicus University in Toruń, Wile\'{n}ska 4, 87-100 Toru\'{n}, Poland}

\affil[3]{Karlsruhe Institute of Technology, Institute of Theoretical Solid State Physics, Kaiserstr. 12, 76131 Karlsruhe, Germany}

\affil[4]{Centro de Física de Materiales, CFM-MPC CSIC-UPV/EHU, Paseo Manuel Lardizabal 5, Donostia-San Sebastián 20018, Spain}

\affil[5]{Department of Applied Physics, School of Architecture, UPV/EHU, Donostia-San Sebastián 20018, Spain}

\affil[6]{Joint Quantum Institute, University of Maryland and National Institute of Standards and Technology, College Park 20742, MD, USA}

\affil[7]{Nanoscale Device Characterization Division, National Institute of Standards and Technology, Gaithersburg 20899, MD, USA}

\affil[8]{Institute of Nanotechnology, Karlsruhe Institute of Technology,  Kaiserstr. 12, 76131 Karlsruhe, Germany}

\affil[9]{Donostia International Physics Center (DIPC), Paseo Manuel Lardizabal 4, Donostia-San Sebastián 20018, Spain}

\date{} 

\begin{document}
\twocolumn[
  \begin{@twocolumnfalse}
    \maketitle
    \begin{abstract}
      We investigate the spontaneous emission of a quantum emitter placed near finite atomic chains described by the Su-Schrieffer-Heeger model. By tuning the chain from a metallic to a semiconducting character through dimerization, we analyze how the electronic structure of an atomic-scale scatterer, together with the illumination profile, determines the strength, bandwidth, and spatial structure of the spontaneous emission enhancement. The resonant enhancement depends on both spectral matching and the emitter position relative to the induced charge and field patterns. In metallic chains, the nearly harmonic energy spectrum leads to overlapping optical transitions and strong resonances. For semiconducting chains, dimerization opens a gap and leads to a modified density of electronic states, redistributing the response over a broader spectral range and reducing its peak intensity. 
Coulomb interactions affect the optical resonances through blueshift or modification of their spatial character, and act differently on collective and single-particle-like excitations.  
Finally, we show that the same mechanisms demonstrated in atomic chains persist in realistic quasi-one-dimensional systems, including trans-polyacetylene chains and graphene nanoribbons.
Our findings indicate a general direction for engineering light–matter interactions at the atomic scale, and may offer design principles for tunable nanoscale photonic and quantum-optical devices.
    \end{abstract}
    \medskip
    {\small\noindent\textbf{Keywords:} spontaneous emission, Purcell effect, Su--Schrieffer--Heeger (SSH) model, atomic chains, graphene nanoribbons\par}
    \vspace{1em}
  \end{@twocolumnfalse}
]
\saythanks
\hypersetup{pdfkeywords={spontaneous emission, Purcell effect, Su-Schrieffer-Heeger (SSH) model, atomic chains, graphene nanoribbons}}

Controlling the spontaneous emission of quantum emitters using nanostructures is one of the main goals of quantum nanophotonics \cite{gonzalez2024}. In conventional plasmonic and dielectric systems, emission enhancement is commonly understood in terms of optical resonances whose characteristics are determined by the local density of electromagnetic states \cite{Novotny_Hecht_2012}. At the atomic scale, however, the origin of these resonances becomes less clear. Finite structures composed of a few tens or hundreds of atoms retain both the discrete nature of individual electronic states and the collective response associated with many-electron interactions. As a result, their optical spectra may contain resonances of different physical character \cite{bernadotte2013,townsend2014,bursi2016,zhang2017,mueller2020, yan2007}.

Recent studies have shown that resonances in nanoscale systems can range from predominantly single-particle excitations to genuinely collective modes \cite{townsend2014,zhang2017,mueller2020,bryant2016}. 
Several approaches have been proposed to distinguish between these limits, including charge-distribution-based measures such as the plasmonicity index \cite{bursi2016} and the generalized plasmonicity index \cite{zhang2017}, as well as energy-space approaches that identify the role of electron-electron interactions in the excitation mechanism \cite{bernadotte2013,townsend2014,mueller2020,mueller2021}. While these methods provide insight into the nature of the resonances themselves, much less is known about how the resonance character affects the spontaneous emission of a nearby quantum emitter.

The spontaneous emission of a quantum emitter depends on the modes supported by the nanostructure and their character, and on how efficiently these modes couple to external electromagnetic field \cite{gonzalez2024, Novotny_Hecht_2012}. The polarization, spatial profile, and degree of inhomogeneity of the excitation field determine the optical selection rules and transition moments involved in the light–matter coupling. A spatially homogeneous field primarily probes dipolar transitions, whereas the strongly inhomogeneous near field of a localized dipole can relax symmetry-based selection rules and couple to higher-order transition moments \cite{rusak2019, chen2022, gonccalves2020}. Thus, the excitation field provides an additional control knob over which resonances contribute strongly to the optical response.

One-dimensional atomic chains are a convenient platform for addressing this question since they sustain features of both metallic and semiconducting systems \cite{Heeger1988,roth1987}. Their electronic structure can be continuously tuned from metallic to semiconducting by introducing dimerization, as described by the Su-Schrieffer-Heeger (SSH) model \cite{su1979, Heeger1988, vega2020, Mueller2021chain}. This modification opens an energy gap around the Fermi level, changing the distribution of electronic states and the optical resonances supported by the structure. The chains, therefore, offer a basic setting in which the relation between electronic structure, resonance character, and spontaneous-emission enhancement can be examined systematically.

In this work, we investigate the spontaneous emission of a quantum emitter placed close to finite SSH chains \cite{su1979}. We vary the emitter position, the emitter-chain separation, the degree of dimerization, and the strength of Coulomb interactions to identify the mechanisms governing emission enhancement at the atomic scale. We first examine the resonant and spatial character of the enhancement in linear chains, and then compare it with the response of dimerized chains, where the strength and spectral distribution of the emission characteristics are determined by the energy anharmonicity and the modified density of electronic states.
We characterize how the structure of the chain and of the illuminating field impact the emission pattern of the quantum emitter.
We further characterize the role of Coulomb interactions on the width and strength of single-particle and collective resonances. 
Our observations obtained for the SSH model explain the optical response of realistic systems such as trans-polyacetylene chains and graphene nanoribbons.
The systems investigated in this manuscript are also relevant to scanning near-field optical microscopy (SNOM), especially scattering-type microscopy, in which an atomic force microscope tip probes the local optical near field of a nanostructure \cite{keilmann2004,fei2012}. Similar to the spontaneous emission rate of a quantum emitter, the detected near-field signal is governed by the local electromagnetic response. Thus, near-field characteristics investigated in this work are important for interpreting both emitter characteristics and tip-based optical nanoscopy.

\begin{figure}[!tbh]
    \centering
    \includegraphics[width=0.49\textwidth]{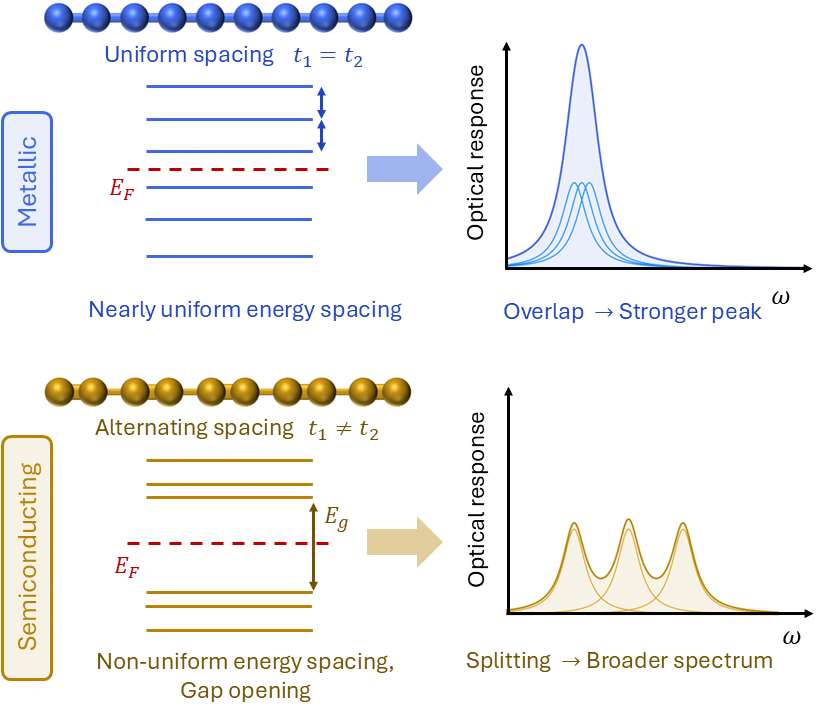}
    \caption{Scheme of SSH chains investigated in this work: metallic (blue) and semiconducting (golden). Left: spatial arrangement of atoms and energy level distribution around the Fermi energy. Right: resulting emission spectrum.}
    \label{fig:schemes}
\end{figure}

\section{Model}
\begin{figure*}[!t]
    \centering
    \includegraphics[width=0.95\textwidth]{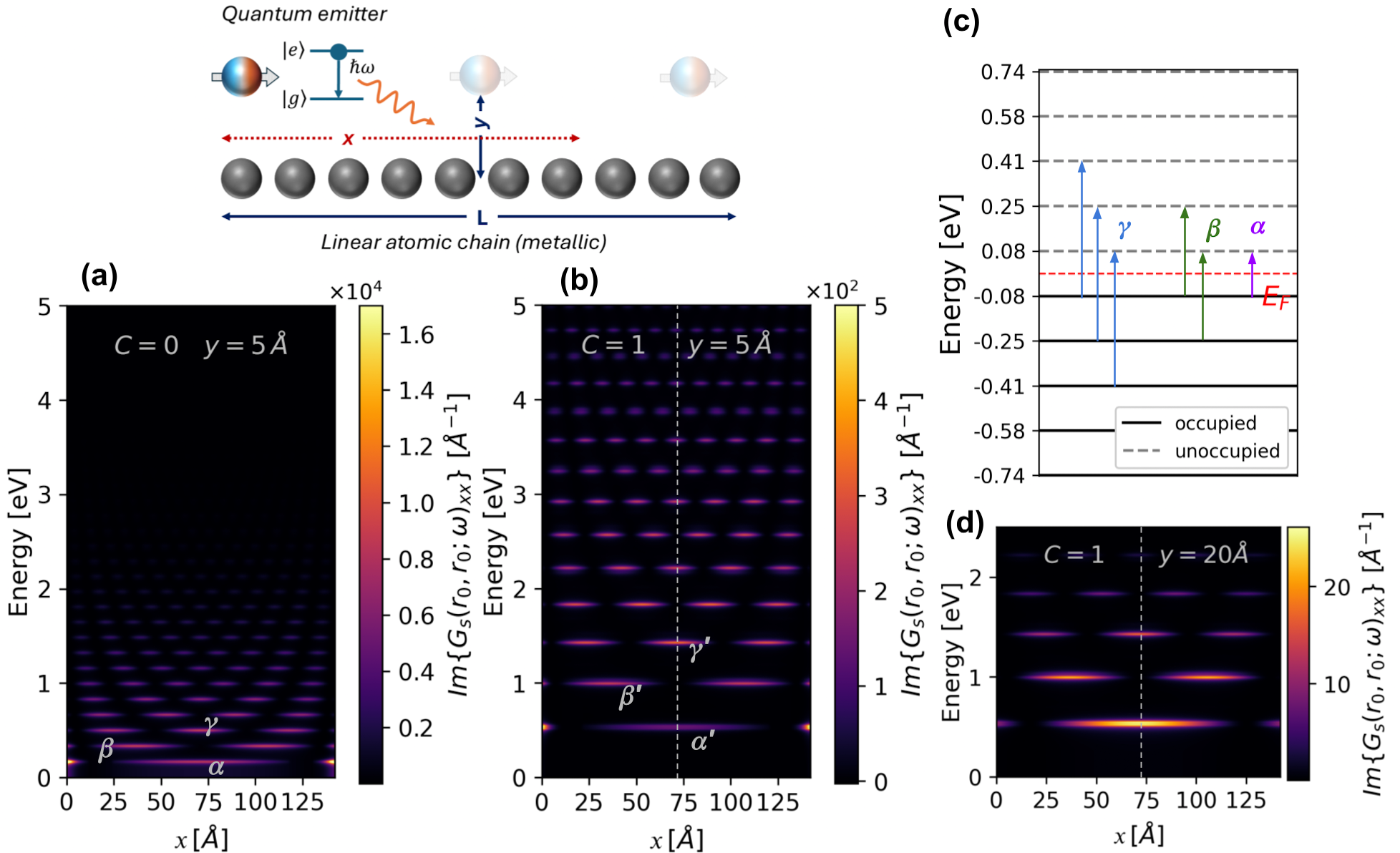}
    \caption{
    Imaginary part of the scattered Green's function, $\mathrm{Im}\{\vec{G}_s(\vec{r}_0,\vec{r}_0;\omega)_{xx}\}$, for a 100-atom metallic chain as a function of the emitter position along the $x$-direction, for different Coulomb scaling parameters $C$ and chain-emitter separations $y$:  
    (a) $C = 0$ and $y = 5~\text{\AA}$;  
    (b) $C = 1$ and $y = 5~\text{\AA}$;  
    (d) $C = 1$ and $y = 20~\text{\AA}$.  
    The first three modes in (a) are labeled $\alpha,\,\beta,$ and $\gamma$, and their Coulomb-interacting counterparts in (b) are labeled $\alpha',\,\beta',$ and $\gamma'$.  
    The dashed white vertical line in (b) and (d) marks the response when the emitter is positioned at the center of the chain.  
    (c) Energy levels near the Fermi level (red dashed line) and the eigenstates involved in the transitions corresponding to modes $\alpha$, $\beta$, and $\gamma$ are indicated.
        }
    \label{figs:LC_xx}
\end{figure*}

The SSH model \cite{Heeger1988}, formulated within the tight-binding framework, provides a fundamental description of the electronic and topological properties of polyene and polyacetylene molecules, graphene nanoribbons, as well as one-dimensional atomic chains in general.
Despite its conceptual simplicity, the model serves as a powerful tool for exploring the underlying physics of such one-dimensional systems.
When we vary the interatomic spacing, the hopping amplitude is modulated, and the system exhibits distinct electronic phases, including the metallic and semiconducting phases (Fig.~\ref{fig:schemes}).
We consider a quantum emitter modelled as an optically coupled two-level adatom with transition energy $\hbar\omega$ positioned in proximity to an atomic chain (see sketch in Fig.~\ref{figs:LC_xx}) and analyze its spontaneous emission rate depending on the nature of the atomic chain, which can be metallic or semiconducting.

The total spontaneous emission rate of the quantum emitter $\Gamma$ is modified in the presence of a scatterer, modelled here as the SSH chain. The total spontaneous emission rate is given by \cite{Novotny_Hecht_2012}
\begin{gather}
    \label{eq:tot_gamma}
   \Gamma = \Gamma_0 + \Gamma_s, \nonumber\\
   \mathrm{where} \quad\Gamma_s= \frac{2\omega^2 }{\hbar \varepsilon_0 c^2} 
\left[ \vec{\mu}_{eg} \cdot \operatorname{Im} \left\{ G_s(\vec{r_0}, \vec{r_0}; \omega) \right\} \cdot \vec{\mu}_{eg} \right].
\end{gather}
Here, $\Gamma_0=\frac{\omega^3 |\vec{\mu}_{eg}|^2}{3\pi\varepsilon_0\hbar c^3}$ and $\Gamma_s$ are the contributions arising from coupling of the emitter to the free space via the Weisskopf–Wigner rate \cite{weisskopf1930} and to the field induced by the scatterer, respectively. 
The key quantity that determines the emission rate in Eq.~\ref{eq:tot_gamma} is the imaginary component of the scattered part of the dyadic Green's tensor \cite{Novotny_Hecht_2012} $ {G_s}(\vec{r}=\vec{r_0},\vec{r'}=\vec{r_0}; \omega)$ evaluated at the position $\vec r_0$ of the quantum emitter. Here, $\vec r$ and $\vec r'$ denote the observation and source coordinates, respectively, which coincide in this case.
The dyadic Green's tensor relates the field at point $\vec{r}$ scattered by the atomic chain (scatterer) with a point 
dipole (source) with transition dipole moment $\vec{\mu_{eg}}$, located at $\vec r'$ emitting at frequency $\omega$.
In Eq.~\ref{eq:tot_gamma}, the Green's tensor is evaluated for the dipole source $\vec{\mu_{eg}}$ representing the transition dipole moment between two states of the quantum emitter separated by energy $\hbar\omega$.
Finally, $\hbar$, $\varepsilon_0$, and $c$ are the reduced Planck constant, the vacuum dielectric permittivity, and the speed of light in vacuum.

 The spontaneous emission rate $\Gamma$ can be tuned by engineering the geometry of the scatterer, as reflected in the Green's tensor. In this work, the chain geometry, representative of quasi-one-dimensional materials in general, is described by a tight-binding (TB) Hamiltonian with a single orbital per chain site \cite{Heeger1988,su1979}, 
\begin{equation}
    \label{eq:tb_hamiltonian}
    H_\mathrm{TB}^\mathrm{chain}=\sum_{l=1}^{N-1} t(1-(-1)^{l}\Delta) (|l\rangle \langle l+1| + \mathrm{H.c.}).
\end{equation}
Here, $|l\rangle$ represents the single-electron orbital localized in the vicinity of the host atom at position $\vec{r_l}$, and $N$ denotes the number of atoms in the chain. The parameter $t$ describes the TB nearest-neighbor hopping amplitude between adjacent orbitals $l$ and $l'$. We adopt a value of $t = -2.66$ eV, which corresponds to the hopping energy of an electron between neighboring carbon $p_z$ orbitals within the standard $\pi$-orbital description of graphene \cite{reich2002}, close to the commonly used value in the SSH model of polyacetylene chains \cite{su1979} and typical for graphene nanoflakes \cite{CastroNeto2009}, to be considered later in this work. The symbol
$\Delta$ represents the dimerization parameter, where $\Delta = 0$ describes a metallic, and $\Delta > 0$ a semiconducting chain. \footnote{
The investigated chains are finite structures with discrete energy levels, however, the strictly metallic behavior occurs in the infinite-chain limit.
}
The abbreviation H.c. stands for the Hermitian conjugate. 
The energy spectra for both types of chains are shown in Fig.~\ref{figs:LC_xx}(c) and Fig.~\ref{figs:ImG_dc_Xx}(a). 
The HOMO-LUMO gap increases due to the $\Delta$ parameter in the dimer chain. 

The ground state of the $N_e$-electron system is constructed from the single-particle eigenstates of the TB Hamiltonian according to the Aufbau principle, which yields the initial density matrix  $\rho(0) =\frac{2}{N_e}\sum_{j=1}^{N_e/2}|j\rangle \langle j|$ \cite{cox2014, kosik2022}, which corresponds to a uniform charge distribution. Note that, in this work, we set $N_e=N$ and select $N$ to be even.
The factor of 2 accounts for spin degeneracy, ensuring normalization $\mathrm \, {Tr}[\rho(0)]=1$.

The atomic chain is under illumination by a classical dipole source, which represents the quantum emitter transition. The interaction is obtained by evaluating the external dipolar potential at each atomic site $\vec r_l$. Thus, the light-matter interaction Hamiltonian at time $\tau$ is given by \cite{Mueller2021chain,vega2020},
\begin{equation}
    \label{eq:int_hamiltonian}
    H^\mathrm{int}=-e\sum_{l=1}^{N} \Phi^{\mathrm{dip}}(\vec r_l,\tau)\ket{l}\bra{l},
\end{equation}
where
\begin{equation}
    \Phi^{\mathrm{dip}}(\vec r,\tau) =\frac{1}{4\pi\varepsilon_0}\frac{\vec\mu(\tau)\cdot(\vec r-\vec{r}')}{|\vec r-\vec r'|^3},
\end{equation}
is the near-field dipolar potential at position $\vec r$ due to a point dipole with dipole moment $\vec\mu(\tau)$ located at $\vec r'$, and $e$ is electronic charge.

This external driving field redistributes the ground-state charge, giving rise to a time-dependent Coulomb interaction, expressed as \cite{cox2014, kosik2022},
\begin{equation}
    \label{eq:ind_hamiltonian}
    H^{\mathrm{ind}} = C  \sum_{ll'} \sum_{l''} v_{ll''}\, \rho^{\mathrm{ind}}_{l''l''}(\tau)\, \delta_{ll'}\, |l\rangle \langle l'| ,
\end{equation}
where $C \in [0, 1]$ is the Coulomb scaling parameter, with $C=1$ representing a realistic interacting system within the scope of this model. The matrix element $v_{ll'}$ is the Coulomb interaction \cite{cox2014, kosik2022} between the sites $l$ and $l'$, while the induced charge at a particular site $l$ is given by $\rho_{ll}^{\text{ind}}(\tau) = -eN_e(\rho_{ll} (\tau) - \rho_{ll}(0))$.

The system dynamics are then obtained by propagating the density matrix according to the master equation \cite{breuer2002}
\begin{equation}
    \label{eq:master_eqn}
    \frac{\partial }{\partial \tau}\rho (\tau)=-\frac{i}{\hbar}[H^\mathrm{total}(\tau),\rho (\tau)]-\frac{1}{\tau_\mathrm{rel}}(\rho(\tau)-\rho(0)),
\end{equation}
with $H^\mathrm{total}=H_\mathrm{TB}^\mathrm{chain}+H^\mathrm{int} + H^\mathrm{ind}.$
The second term on the right-hand-side of Eq.~\ref{eq:master_eqn} phenomenologically accounts for relaxations through the coherence time $\tau_\mathrm{rel}$, whose inverse determines the linewidth of the optical resonances discussed below. A constant value $\tau_\mathrm{rel} = 40\,\hbar/\mathrm{eV}$ is used for all structures unless stated otherwise, chosen to be comparable to the coherence times (20 fs or less) reported in Ref.~\cite{yan2013}.
We have assumed here that the system relaxes to the ground state $\rho(0)$. 

From the resulting charge dynamics, the scattered near field is calculated from the dominant induced charge density contribution as \cite{cox2014}
\begin{equation}  
\vec{E}^{\text{scat}}(\vec{r},\tau) \simeq \frac{1}{4\pi \varepsilon_0} \sum_{l=1}^{N}\rho^{\text{ind}}_{ll}(\tau)\frac{(\vec{r_l}-\vec{r})}{|\vec{r_l}-\vec{r}|^3}. 
\end{equation}
Subsequently, the corresponding elements of the Green’s tensor are obtained as
\begin{equation}
    \label{eq:Gs}
{G_{s}}(\vec{r}_0,\vec{r}_0;\omega)_{ij} = \frac{\varepsilon_0c^2}{\omega^2} \frac{\tilde E_i^{\mathrm{scat}}(\vec{r}_0,\omega)}{\tilde\mu_j(\omega)},
\end{equation}
where the quantities with the tilde symbol denote frequency-dependent vectors obtained through Fourier transformation. Based on this Green's function, we investigate the impact of chain dimerization and the spatial profile and symmetry of the incident field on the emission rate, following Eq.~\ref{eq:tot_gamma}.

\section{Results}

In this section, the optical properties of emitters near linear and dimerized atomic chains are examined as functions of their separation from the chain and position along the chain.
The optical response is found to differ qualitatively between metallic and semiconducting chains: metallic chains exhibit approximately equally spaced energy levels near the Fermi level, which leads to strongly overlapping resonances. In contrast, semiconducting chains have an energy gap, observed to result in an anharmonic spectrum and more complex optical response (see Fig.~\ref{fig:schemes}). In the following, we analyze the consequences of neighboring quantum emitters on spontaneous emission enhancement. 

\begin{figure*}[!tbh]
    \centering
    \includegraphics[width=.86\textwidth]{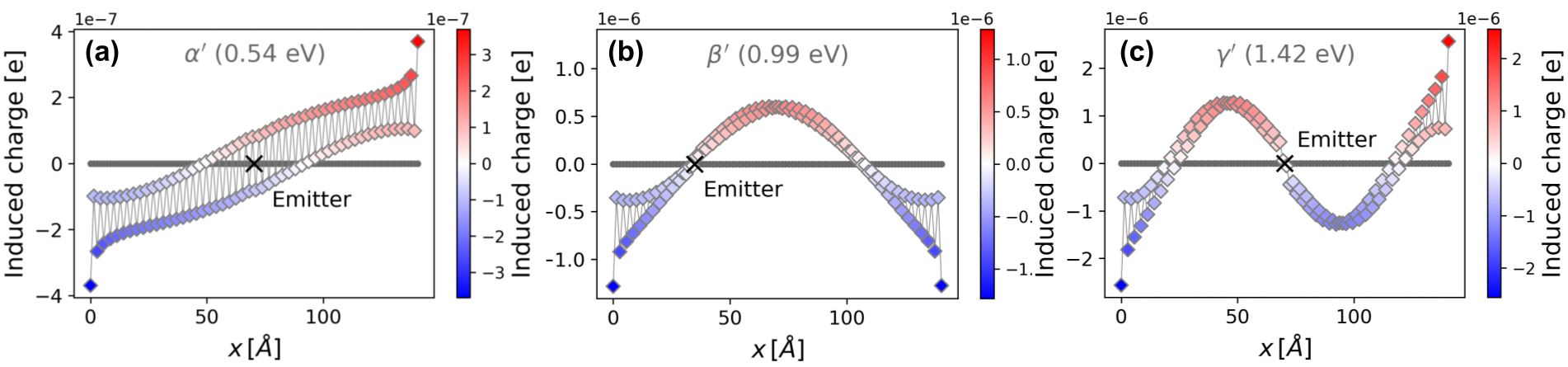}
    \caption{
   Snapshots of the time-dependent induced charge distribution on the linear chain, taken at the turning points of charge oscillations, for the resonances  
    (a) $\alpha'$ ($0.54\, \mathrm{eV}$),  
    (b) $\beta'$ ($0.99 \,\mathrm{eV}$), and  
    (c) $\gamma'$ ($1.42 \,\mathrm{eV}$), respectively.  
    The quantum emitter is located at the center of the chain in (a) and (c), and at the quarter position in (b).  
    The black cross indicates the $x$ coordinate of the emitter. In all cases, the emitter–chain separation is fixed at $y = 5~\text{\AA}$.  
    For each panel, the left $y$-axis and the colour bar represent the induced charge.
    }
    \label{figs:IC_Xx_all}
\end{figure*}

\subsection{Metallic chain}
\subsubsection{Spectral response as a function of emitter position along the chain}
\label{sec:x-dependence}
We start with a quantum emitter optically coupled to a metallic chain of $ N=100$ atoms, with the length of $L=(N-1)a$, where $a=1.42$~\AA\ is the interatomic distance.
The quantum emitter is moved along the length of the chain ($x$-axis), maintaining a separation from the chain of $y=5$ or $20$\ ~\AA. The emitter is modeled as a point dipole emitting an ultrashort pulse at time $\tau=0$, characterized by a sufficiently broad spectral bandwidth. Unless stated otherwise, the dipole is oriented along the $x$ direction for most efficient coupling. The dependence on other dipole orientations is examined in Appendix~\ref{appendix:rotation}. 
In Fig.~\ref{figs:LC_xx}(a), (b), (d), each point on the horizontal axis corresponds to a specific emitter-chain coupling position along the chain.
For each point, an independent simulation is carried out to compute the frequency-resolved (vertical axis) Fourier component of the scattered field $\tilde{E}_x^{\mathrm{scat}}(\vec{r}_0, \omega)$ along $x$-direction, from which $\mathrm{Im}\{\,G_s(\vec{r}_0,\vec{r}_0;\omega)_{xx}\}$ is calculated. For brevity, from now on we denote $\mathrm{Im}\{\,G_s(\vec{r}_0,\vec{r}_0;\omega)_{xx}\}\equiv\mathrm{Im}\,G_s$.
\footnote{For better visibility of higher and lower order modes together and comparisons across systems of different sizes we plot higher-order roots of $\mathrm{Im}G_s$ or related quatities.}

The spontaneous emission of the emitter is enhanced at the chain resonance frequencies. 
For a vanishing Coulomb interaction strength, i.e., $C=0$, in Fig.~\ref{figs:LC_xx}(a), each resonance can be linked to a transition between a pair of eigenstates (as indicated with $\alpha,\beta,$ and $\gamma$ in panels (a) and (c)).  
Shifting the emitter along the chain reveals the spatial structure of the modes, in analogy to how they would be scanned in a SNOM experiment \cite{keilmann2004,fei2012}. These modes show increasing resonance order as the energy grows: the first-order mode ($\alpha$, lowest in energy) features a single bright spectral region, the second-order mode ($\beta$, second lowest energy) two bright regions, and so forth. The positions of the regions of bright spectral response indicate, respectively, the frequencies (vertical axis) and chain locations (horizontal axis) where the quantum emitter couples best. Figure~\ref{figs:LC_xx}(b) shows the corresponding spectra with the full Coulomb interaction ($C=1$). For metallic chains, the impact of Coulomb interactions is limited to a substantial blueshift of the modes. In Section~\ref{sec:Coulomb}, this result will be compared with the case of semiconducting chains, where the Coulomb impact becomes considerably richer. 

\begin{figure*}[t]
    \centering
    \includegraphics[width=0.84\textwidth]{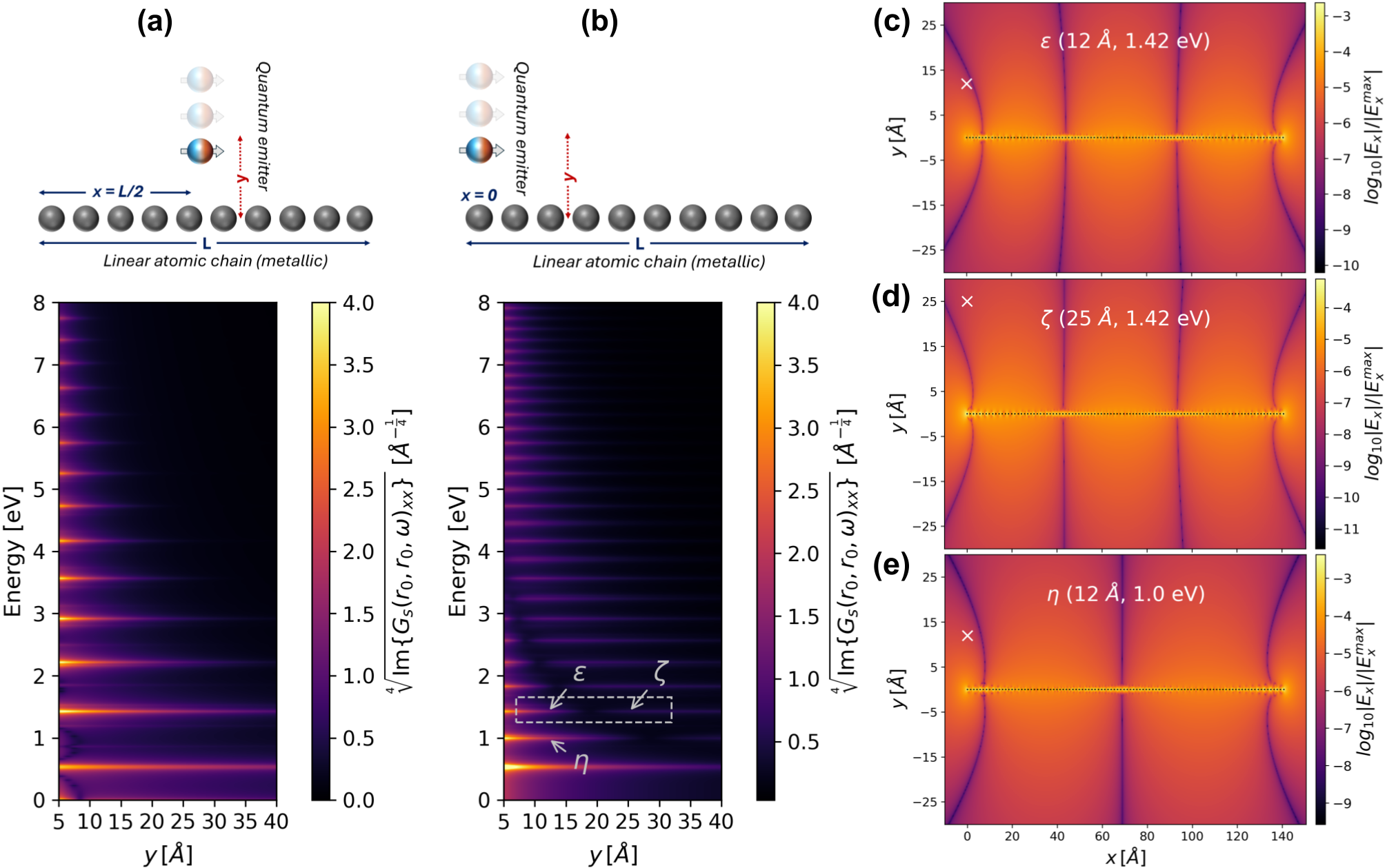}
    \caption{ Fourth root of $\mathrm{Im}\{\vec{G}_s(\vec{r}_0,\vec{r}_0;\omega)_{xx}\}$ for a 100-atom linear chain plotted as a function of the chain–emitter distance $y$: emitter positioned above the (a) center $(L/2)$ and (b) edge of the chain. 
    The third-order mode is highlighted with a dashed rectangle in panel (b). The points labeled $\varepsilon$ and $\zeta$ correspond to $y = 12~\text{\AA}$ and $y = 25~\text{\AA}$, respectively, on the third-order mode, representing positions before and after the sudden dip in the mode near $y = 20~\text{\AA}$. The point labeled $\eta$ lies on the second-order mode at $y = 12~\text{\AA}$.
    (c)--(e) Normalized scattered field distributions around the chain corresponding to the points $\varepsilon$, $\zeta$, and $\eta$.  
    In each panel, the white cross marks the position of the quantum emitter.
    }
    \label{figs:L100_y_dependence}
\end{figure*}

In the non-interacting case ($C=0$), the origin of these modes can be understood from the energy spectrum of the linear chain near the Fermi level [Fig.~\ref{figs:LC_xx}(c)]. The first-order mode originates from the HOMO-LUMO transition $\alpha$, whereas the higher-order modes involve higher-order transitions. For sufficiently long linear chains, the levels near the Fermi energy become nearly equidistant. Because such a level structure is characteristic of a harmonic oscillator, we refer to the spectrum of the linear chain in this regime as harmonic. As a result, the corresponding low-energy transitions occur in closely spaced spectral ranges, and their resonances overlap in the $\mathrm{Im} G_s$ spectra. The overlap reinforces their contributions at similar frequencies, giving rise to a strong optical response. 

The quantum emitter, acting as a source of driving dipolar illumination, excites several charge modes induced in the chain. When moved along the chain, it simultaneously acts as a detector, profiling the underlying induced charge distribution in the chain.
Fig.~\ref{figs:IC_Xx_all}(a)--(c) shows the induced charge distributions for the first three orders of modes $\alpha',\ \beta',$ and $\gamma'$ marked in Fig.~\ref{figs:LC_xx}(b). 
In the $y=5$~\AA\ configuration, where the emitter is positioned very close to the chain, the interaction occurs with the local induced dipole moment at the nearby sites.
For the first-order mode [Fig.~\ref{figs:IC_Xx_all}(a)] the envelope of the induced charge distribution has a node at the center of the chain, which carries the strongest local induced dipole moment, while for the $n$th-order mode there are $n$ nodes. 
Coupling the emitter very close to this site gives rise to a strong dipole-dipole interaction with the chain, resulting in a bright response at the center of the chain in the $\mathrm{Im}\, G_s$ spectra (Fig.~\ref{figs:LC_xx}).
For higher-order modes, the strong response in the $\mathrm{Im}\, G_s$ spectra maps directly onto the interaction with the local charge nodes of the corresponding induced charge distributions.
For each mode, the regions of a bright response are separated by narrow dark ones.
These correspond to the extrema of the induced charge distribution, where charge variation between adjacent sites is minimal and the local induced dipole moment is negligible.
Note that for low-order modes, a strong response at the chain edges can be observed, originating from the finite-size effect. The dark regions in the $\mathrm{Im}\, G_s$ spectra can also be understood through nodal planes of the electric field induced at the chain [Fig.~\ref{figs:L100_y_dependence}(c)--(e)]: When the emitter lies on the nodal plane, it experiences no induced field, which results in a dark region in the spectrum and lack of spontaneous emission enhancement.

In the $y = 20$ ~\AA\ configuration [Fig.~\ref{figs:L100_y_dependence}(d)], the emitter is more distant from the chain and, therefore, couples to the global charge distribution rather than probing local features. As a result, the fine structure of the modes is no longer detectable, and only the lower-order modes remain visible in the spectra [Fig.~\ref{figs:LC_xx}(d)]. 
Thus, for large distances, emission from quantum emitters resonant with the lower-order modes of the chain will be efficiently enhanced.

\begin{figure*}[t]
    \centering
    \includegraphics[width=0.88\textwidth]{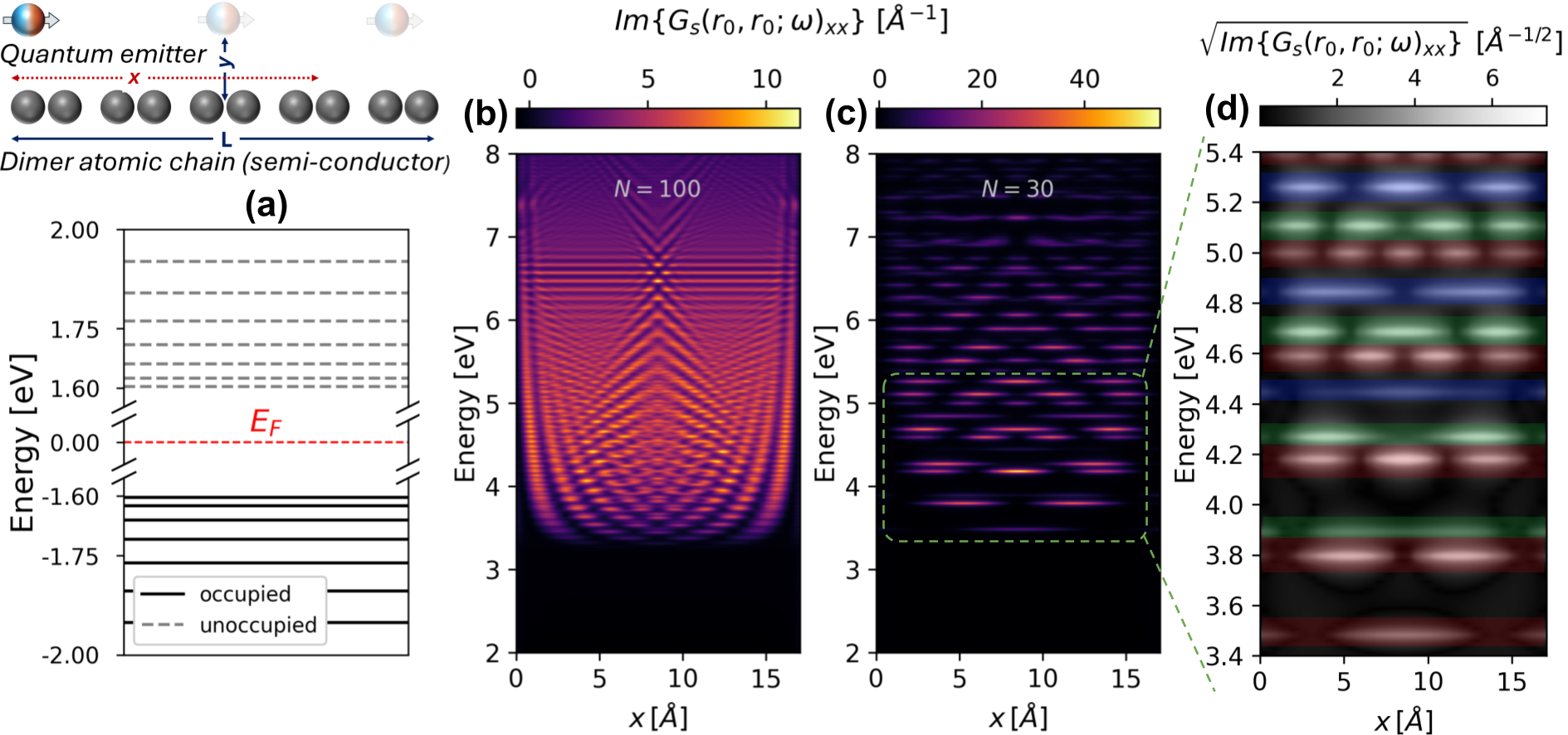}
    \caption{
    (a) Energy levels of an $N=100$ atom semiconducting chain ($\Delta = 0.3$) near the energy gap induced by dimerization.
    (b) $\mathrm{Im}\{\vec{G}_s(\vec{r}_0,\vec{r}_0;\omega)_{xx}\}$ spectra for 
    the same chain as a function of the emitter position along the $x$-direction with Coulomb interaction included, and  
    (c) an $N=30$ atom semiconducting chain under the same conditions.  
    (d) Enlarged view of subfigure (c) with square root of $\mathrm{Im}\{\vec{G}_s(\vec{r}_0,\vec{r}_0;\omega)_{xx}\}$ for better visibility, highlighting the first three linear-chain-like spectral patterns in red, green and blue colours.  
    Please note that the energy scale begins at $2$ eV.
            }
    \label{figs:ImG_dc_Xx}
\end{figure*}

\subsubsection{Spectral response as a function of emitter-chain distance}
We now investigate the dependence of the spectral response of the chain on the distance of the quantum emitter in the interacting case ($C=1$). 
The quantum emitter is first placed in two configurations: (a) above the middle of the chain and (b) above the first atom of the chain.
To investigate the spectral response dependence on the emitter-chain distance, the emitter is displaced along the $y$-direction from $y=5$ to $y=40$~\AA\ (horizontal axis in Fig. \ref{figs:L100_y_dependence}(a), (b)).
At each $y$ position, the imaginary part of the scattered Green’s function is evaluated as in the previous section.

For both configurations, increasing the emitter–chain distance leads to a rapid decrease in the intensity of higher-order modes in the $\mathrm{Im}\, G_s$ spectra, because of the inverse power-law decay of the scattered field generated by the underlying multipolar charge distribution \cite{griffiths2023}. 

For the emitter positioned in the middle of the chain, even-order modes are quenched [Fig.~\ref{figs:L100_y_dependence}(a)].
This effect is related to the selection rule forbidding dipole transitions between eigenstates of the same parity in a system driven by a field symmetric with respect to a mirror plane passing through the center of the chain \cite{sakurai1994}. This mirror symmetry of the field arises when the emitter is positioned at the center of the chain (white dashed line in Fig.~\ref{figs:LC_xx}(b)). The same quenching effect is visible in Fig.~\ref{figs:LC_xx}(a) and (b) for the emitter position above the middle of the chain. 
For other emitter positions, the strongly inhomogeneous dipolar near field of the emitter relaxes this selection rule and allows transitions between states of the same parity.
This symmetry aspect is explored further in Section~\ref{subsec:disappearence_absorption_peak} by comparing the absorption spectra and $\mathrm{Im}\, G_s$ spectra in metallic and semiconducting chains.

In the edge configuration [Fig.~\ref{figs:L100_y_dependence}(b)], the power-law decay is superimposed with the dark regions of suppressed emission. As before, the dark regions correspond to the nodal planes of the induced field: The induced field patterns for configurations labeled as $\varepsilon$, $\zeta$, and $\eta$ are shown in Fig.~\ref{figs:L100_y_dependence}(c)--(e). 
In these figures, the white cross marks the emitter position. Thus, an emitter positioned near the edge of the chain may cross the curved nodal surfaces while displaced along the $y$-axis, resulting in local suppression of spontaneous emission. 
Therefore, the subtleties of the field patterns near the nanostructure edge may play a particularly significant role, determining the interaction of these nanostructures with quantum emitters \cite{Mueller2021chain}.

\begin{figure*}[!tbh]
    \centering
    \includegraphics[width=0.85\textwidth]{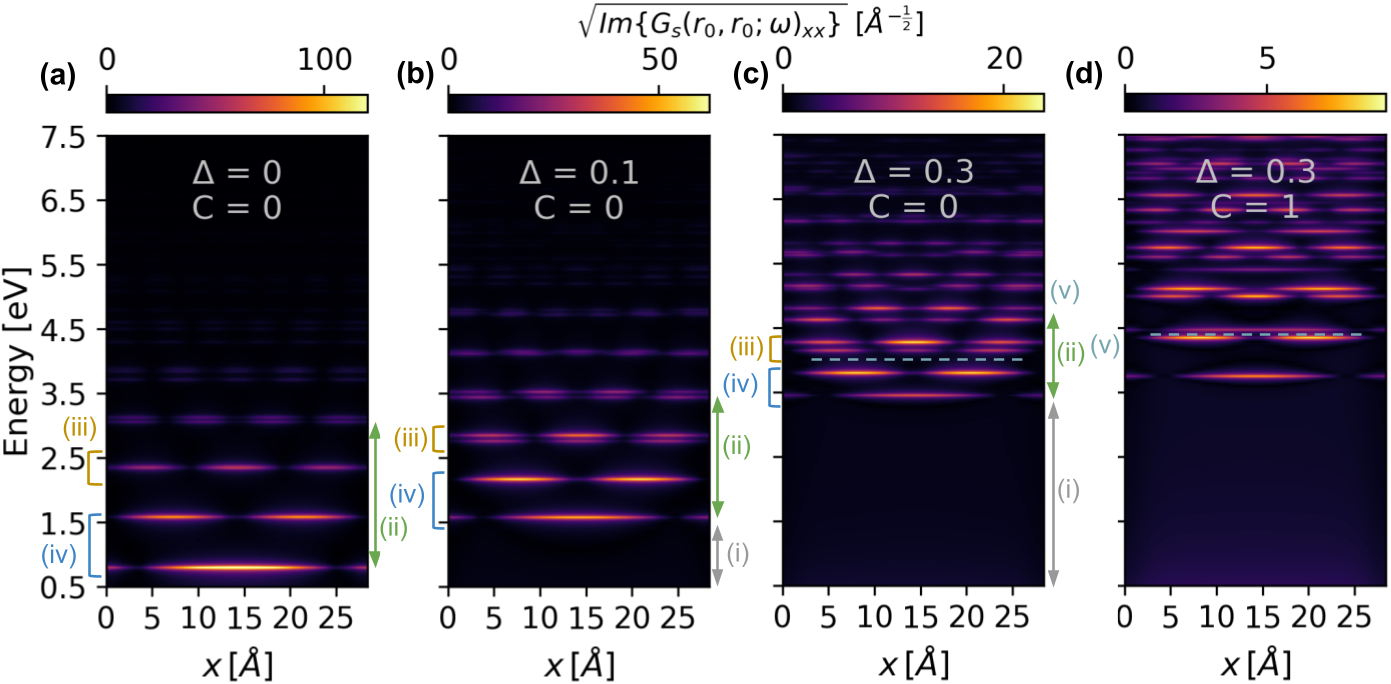}
    \caption{
    Square root of $\mathrm{Im}\{\vec{G}_s(\vec{r}_0,\vec{r}_0;\omega)\}$ spectra for a non-interacting ($C=0$), $N=20$ atom  
    (a) metallic chain ($\Delta = 0$) and semiconducting chains with dimerization parameters    
    (b) $\Delta = 0.1$, and  
    (c) $\Delta = 0.3$.
    (d) interacting semiconducting chain ($C=1,\, \Delta=0.3$).
    Across all panels, several spectral features are highlighted and labeled (i)–(v):
    (i) opening of an energy gap and an overall blueshift of the spectra;  
    (ii) increased density of resonances;  
    (iii) splitting of third-order mode upon dimerization; and  
    (iv) first- and second-order modes remaining unsplit upon dimerization;
    (v) change of spatial symmetry of a mode upon Coulomb interaction. The lower branch of the third-order mode with three bright spots in a non-interacting dimer chain [panel (c)] changes its spatial symmetry profile, having one bright spot in the interacting chain [panel (d)]. }
    \label{figs:delta-dep}
\end{figure*}
\subsection{Semiconducting chain}\label{subsec:semiconducting}
We next analyze dimer chains with an energy gap between the HOMO and LUMO levels. 
We set the dimerization parameter to $\Delta =0.3$ and shift the emitter position along the chain ($x$-axis), while keeping the emitter–chain distance fixed at $y = 5$~\AA.

Figure~\ref{figs:ImG_dc_Xx}(b) and (c) present the $\mathrm{Im}\, G_s$ spectra for interacting ($C=1$) semiconducting chains of $N=100$ and $N=30$ atoms as a function of emitter position along the chain. 
The spectra are significantly more complex than those of the corresponding metallic chain studied in the previous section. The energy levels in the semiconducting chain are no longer equidistant, and the spectrum loses its harmonic character [Fig.~\ref{figs:ImG_dc_Xx}(a)]. The optical resonances undergo a blueshift as the $\Delta$ parameter grows. As the HOMO–LUMO energy gap increases, the transition energies become increasingly non-uniform, which leads to a broader spectral region of enhanced spontaneous emission. The broadening stands in contrast to the distinct resonance peaks observed in the linear chains, as shown for a sufficiently long semiconducting chain in Fig.~\ref{figs:ImG_dc_Xx}(b).
Due to the broadening, the overall spectral intensity of the semiconducting chain is much weaker than that of the metallic chain of the same length, as can be seen by comparing different color-bar scales in Fig.~\ref{figs:ImG_dc_Xx}(b) and Fig.~\ref{figs:LC_xx}(b).

To investigate the origin of the broadening, we study the case of a shorter chain of $N=30$ [Fig.~\ref{figs:ImG_dc_Xx}(c)]. This allows us to attribute the broadening to a splitting of resonances into several spectrally shifted patterns. The overlapping patterns are shown in different colors in the zoomed panel (d), and each of them separately resembles the pattern of a linear chain as in Fig.~\ref{figs:LC_xx}(b). They are, however, blueshifted to a different degree, which leads to a complex spectral pattern. 

For short chains, these patterns are spectrally separated. As the density of electronic states increases with the length of the chain, the number and, hence, the spectral density of optical resonances increase, contributing to the overall broad optical response of the chain.

Fig.~\ref{figs:delta-dep}(a)--(c) shows non-interacting ($C=0$) chains of $N=20$ atoms, with an increasing dimerization parameter $\Delta$, illustrating the evolution of the $\mathrm{Im}G_s$ spectrum as the chain transitions from a metallic to a semiconducting state. The first thing to notice is the clear blueshift of the resonances, related to the opening of the HOMO-LUMO gap [marked as (i) in Fig.~\ref{figs:delta-dep}(b), (c)]. Due to the energy gap opening, the energy levels directly below and above the energy gap become squeezed in the energy space [see Fig.~\ref{figs:ImG_dc_Xx}(a)], which is reflected in the optical spectra as an increased density of resonances above the energy gap (ii). 
In addition, $\Delta > 0$ causes the splittings of the resonances, leading to an even higher density of resonances for order higher than two (marked as (iii) in panels (a)--(c)). For a graphical representation of a modified density of states, see Appendix ~\ref{appendix: splitting}. 

We conclude that, compared to linear chains, dimerized chains exhibit a richer spectral response due to their non-equidistant energy spectrum: The gap introduced by the dimerization parameter induces a blueshift and renders the transition energies increasingly non-uniform as the HOMO–LUMO gap widens. As a result, spontaneous emission enhancement is distributed over a broader spectral range rather than concentrated in sharp resonances. This spectral broadening leads to a significantly reduced overall intensity compared to a linear chain of the same length, which results in weak yet spectrally broad spontaneous emission. 
This conclusion yields
a clear physical interpretation of the mechanisms underlying the broad optical response in nanostructures with a high electronic density of states.

\subsection{Role of illumination symmetry}\label{subsec:disappearence_absorption_peak}

\begin{figure}[!tbh]
    \centering
    \includegraphics[width=0.49\textwidth]{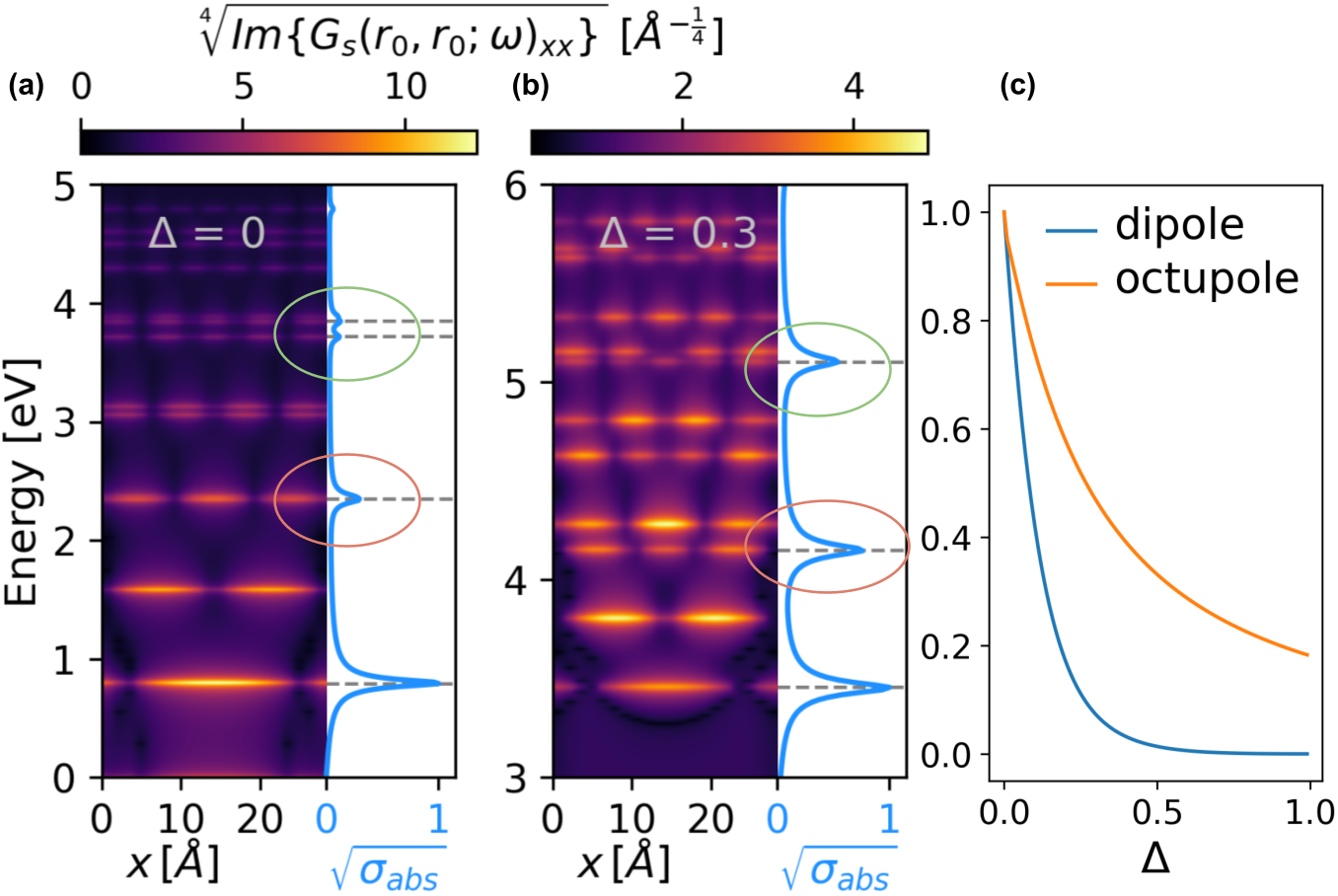}
    \caption{
    Fourth root of $\mathrm{Im}\{\vec{G}_s(\vec{r}_0,\vec{r}_0;\omega)\}$ spectra (left panels, heatmap) as a function of the emitter position along the length of the chain, shown alongside the square root of the corresponding absorption spectra upon plane-wave illumination (right panels, line plots) on a shared energy axis.  
    Results are presented for 
    (a) metallic chain ($\Delta = 0$) and  
    (b) semiconducting chain ($\Delta = 0.3$).
    All even-order modes visible in the $\mathrm{Im}\,G_s$ spectra are absent in the absorption spectra.  
    For the odd-order modes, only a single branch of each multi-branch mode present in the $\mathrm{Im}\,G_s$ spectra survives in the absorption spectra upon dimerization, highlighted by circles.
    (c) Normalized elements of the dipole (blue): $d_{H \to L+2}(\Delta)/d_{H \to L+2}(0)$ and octupolar (orange) moments: $O_{H \to L+2}^{xxx}(\Delta)/O_{H \to L+2}^{xxx}(0)$ between the HOMO and LUMO+2 states of the chain as a function of the dimerization parameter $\Delta$. All panels correspond to 20-atom chains.}
    
    \label{figs:ImG_vs_abs}
\end{figure}

Above, we have studied how the properties of the chain affect the spontaneous emission; we now examine how the optical response is governed by the symmetry of the illumination, selection rules, and oscillator strengths of transitions within the system. In order to investigate the effect of the excitation-field symmetry, we compare the emission and absorption spectra under dipolar and plane-wave illumination, representing spatially inhomogeneous and homogeneous fields, respectively.
In the case of plane-wave illumination, the reflection symmetry is preserved, and transitions between states of the same (even or odd) parity are forbidden by selection rules \cite{sakurai1994}, resulting in the absence of even-order modes in the absorption spectrum (Fig.~\ref{figs:ImG_vs_abs}). 
These even-order modes observed in the spontaneous emission spectra (Figs.~\ref{figs:LC_xx}, \ref{figs:L100_y_dependence}, and \ref{figs:ImG_dc_Xx}) arise from the lifting of dipolar selection rules by the spatially inhomogeneous near field of the dipole, which breaks reflection symmetry. The exception is a dipole positioned above the chain center, which preserves the mirror symmetry and thus blocks transitions between same-parity states (Fig.~\ref{figs:LC_xx}).
All odd-order modes observed in spontaneous emission spectra are also present in absorption spectra for a metallic chain.

\begin{figure*}[!t]
    \centering
    \includegraphics[width=0.95\textwidth]{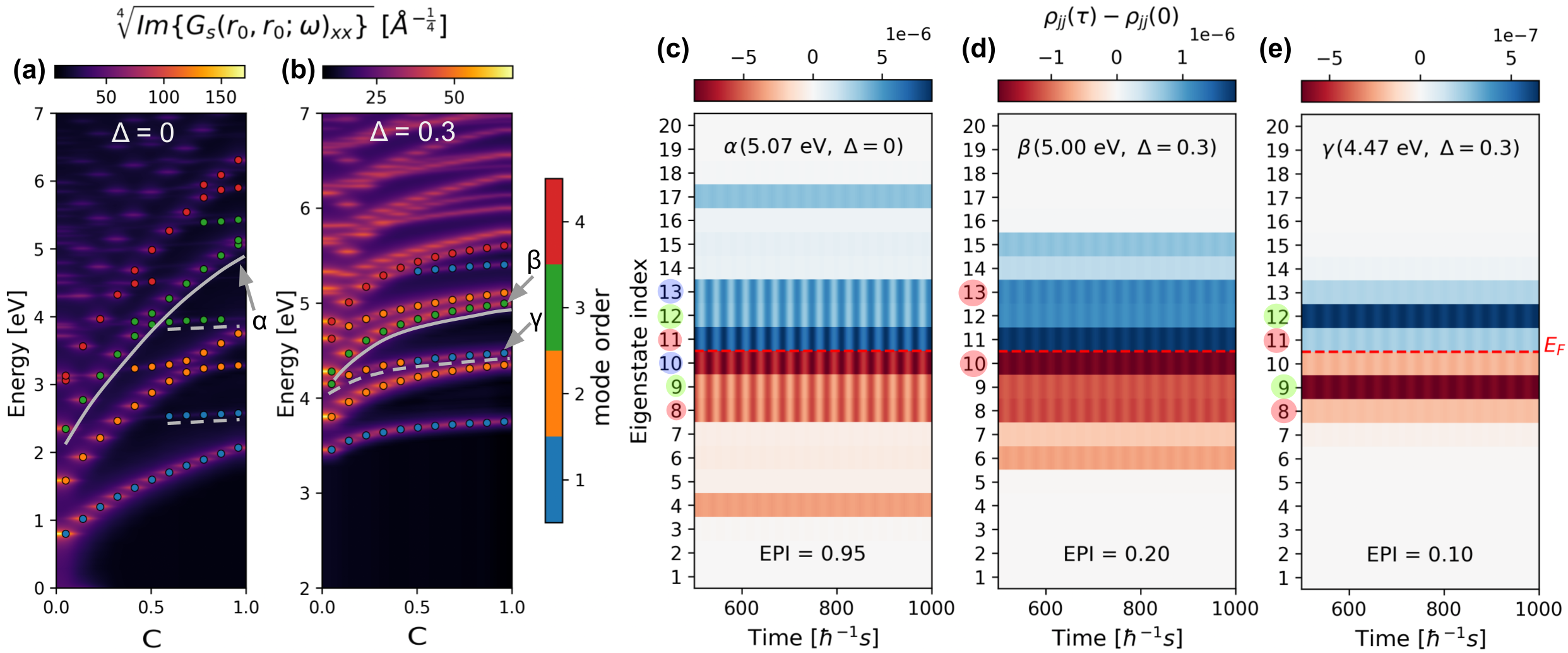}
    \caption{
    (a) Spatially summed fourth root of $\mathrm{Im}\,G_s$ plotted as a function of the Coulomb interaction strength for an $N = 20$ atom metallic chain.  
    Selected modes are tracked across increasing $C$.  
    Colored dots along each tracked branch indicate the mode order (i.e., the number of bright spots in the $\mathrm{Im}\,G_s$ vs.\ $x$ spectra).  
    Solid lines denote modes that undergo strong blueshifts with increasing Coulomb interaction, whereas dashed lines indicate modes that remain nearly flat and weakly dependent on $C$.  
    The point $\alpha$ marks a configuration on the third-order mode at $C = 1$.
    (b) Same analysis for a semiconducting chain with $\Delta = 0.3$.  
    The two branches of the third-order mode ($C = 0$) are tracked with increasing $C$. 
    The points $\beta$ and $\gamma$ indicate representative configurations on these two branches at $C = 1$.
    (c)--(e) Time dynamics of the occupation difference in energy space (relative to the ground-state density matrix) in the light-driven steady state, for excitation by a continuously oscillating dipole in configurations $\alpha$, $\beta$, and $\gamma$, respectively. 
    The color-marked eigenstate indices on the $y$-axis indicate the specific eigenstates that participate in the particular mode for the non-interacting case ($C = 0$), distinguishing the involvement of the same set of states in interacting and non-interacting cases.
    }
    \label{figs:short_chain}
\end{figure*}

Upon dimerization, some odd-order modes present in the absorption spectrum of the metallic chain become quenched in the absorption spectrum, whereas the corresponding modes persist in $\mathrm{Im}G_s$. For example, the upper branches of the third- and fifth-order modes highlighted in Fig.~\ref{figs:ImG_vs_abs}(a), (b) are dominated by transitions between states asymmetrically positioned about the Fermi level. For these transitions, the transition dipole moment decreases rapidly with dimerization [panel (c)]. Since the optical response under homogeneous plane-wave illumination is governed solely by the transition dipole moment, its suppression leads to the disappearance of these modes from the absorption spectrum. In contrast, spatially inhomogeneous dipolar illumination couples to higher-order transition moments in addition to the dipole moment. These higher-order transition moments decrease more slowly with dimerization compared to the dipolar one, allowing the transitions to persist in $\mathrm{Im}G_s$, albeit with reduced strength relative to the metallic chain. For the third-order mode considered here, the longitudinal component of the full octupole moment tensor between states $\ket{i}$ and $\ket{j}$, $O_{ij}^{xxx} \propto\bra{i}x^3\ket{j}$, computed with respect to the center of the chain, is the lowest-order nonvanishing contribution to the oscillator strength for a one-dimensional system, allowing the corresponding modes to persist in $\mathrm{Im}G_s$ [Fig.~\ref{figs:ImG_vs_abs}(c)]. Moreover, these transitions are a finite-size effect; they vanish in an infinite chain as discussed in detail in Appendix~\ref{appendix:transition_moments}.
Transitions between states symmetrically positioned about the Fermi level exhibit a qualitatively different dimerization dependence: their transition dipole and higher-order moments do not vanish with increasing $\Delta$ (Fig.~\ref{figs:TDMs_num}). This behavior predominantly arises from the intrinsic structure of the SSH chain, as discussed in Appendix~\ref{appendix:transition_moments}.

As a complementary classical picture, we consider the spatial Fourier components of the inhomogeneous dipolar field and their overlap with the characteristic wave vectors of the sinusoidal-like charge modes supported by the metallic chain (Fig.~\ref{figs:IC_Xx_all}). Efficient excitation occurs when the field has appreciable Fourier weight at the corresponding mode wave number. Accordingly, the intensity distribution in $\mathrm{Im}G_s$ spectra of a metallic chain reflects the degree of wave-vector matching, which can be tuned through the emitter-chain distance or the chain length. This aspect is discussed in detail in Appendix~\ref{appendix:spatial_mode_matching} for the emitter position above the center of the chain [vertical dashed line in Fig.~\ref{figs:LC_xx}(b), (d)].

The spatial inhomogeneity of the excitation field provides an additional means to control the optical response. While homogeneous illumination probes transitions through their dipole moments, inhomogeneous dipolar illumination accesses the chain's higher-order moments, allowing transitions suppressed by dimerization to remain optically active. Thus, the optical response can be controlled through both the electronic structure of the chain and the spatial profile of the excitation field. At very short emitter–chain distances, electron tunneling may further modify the symmetry and wavefunction parity, providing an additional mechanism for controlling selection rules and transition strengths \cite{Mueller2021chain}.

\subsection{Effect of Coulomb interaction}\label{sec:Coulomb}
We have shown how the distribution of electronic energy levels around the Fermi level determines the strength and spectral distribution of the optical resonances. In metallic chains, the nearly uniform level spacing enables closely spaced transition contributions to reinforce one another, producing strong, well-separated optical resonances. In semiconducting chains, the anharmonic level spacing distributes these contributions over different transition energies, resulting in weaker and more broadly distributed resonances. We now examine how Coulomb interactions modify these responses. Figure~\ref{figs:short_chain} compares short metallic ($\Delta=0$) and semiconducting ($\Delta=0.3$) chains with $N=20$ atoms. The chain length and decoherence time, $\tau_{\mathrm{rel}} = 20\ \hbar$/eV, were chosen so that individual resonances remain distinguishable in the metallic chain despite the Coulomb-induced redistribution of their frequencies.

Increasing the Coulomb interaction blueshifts the resonances in both chain types [Fig.~\ref{figs:short_chain}(a), (b)], with substantially larger shifts in metallic chains. For example, the tracked third-order metallic mode in Fig.~\ref{figs:short_chain}(a) (solid line) shifts significantly by 2.72 eV between $C=0$ and $C=1$. 
Because different resonances undergo different shifts, their spectral separation increases, effectively stretching the spectrum over a wider frequency range.
 
Beyond inducing spectral shifts, Coulomb interactions also modify the spatial profile of the resonances. This effect is more prominent in semiconducting chains [Fig.~\ref{figs:short_chain}(b)]. Several higher-order resonances that are nearly degenerate at $C=0$ separate into branches with different Coulomb-induced blueshifts. For instance, the third-order resonance of the non-interacting ($C=0$) semiconducting chain divides into an upper branch (solid line) that undergoes a pronounced blueshift, and a lower branch that remains weakly dependent on $C$.
This flat lower branch undergoes a change in its spatial symmetry profile, effectively transforming the third-order mode into a first-order mode, as indicated by the change in the number of intensity maxima in the spatially resolved $\mathrm{Im}G_s$ spectra, excluding edge effects. For direct spatial comparison, see the modes marked by dashed underlines in Fig.~\ref{figs:delta-dep}(c) and (d).
We attribute this behavior to a competition between Coulombic forces, favoring charge accumulation at the chain ends, and the semiconducting transition induced by $\Delta$, which favors an insulating pattern.
Coulomb interactions thus act in two ways on the branches produced by dimerization: the interactions blueshift the branches to different degrees and change the spatial symmetry of some of them. Together with the dimerization-induced splitting, this explains why the spectra of interacting semiconducting chains in Fig.~\ref{figs:ImG_dc_Xx}(d) appear as several overlapping, mutually shifted metallic-chain-like patterns.

The collective or single-particle-like character of resonances can be identified in the dynamics of the system described in the energy basis in the driven stationary regime [Fig.~\ref{figs:short_chain}(c)--(e)]. All three examples correspond to $C=1$. Panel (c) shows the occupation difference dynamics $\rho(\tau)-\rho(0)$ on the predominantly collective resonance on a metallic chain, marked by $\alpha$ in panel (a). 
\begin{figure}[!h]
    \centering
    \includegraphics[width=0.48\textwidth]{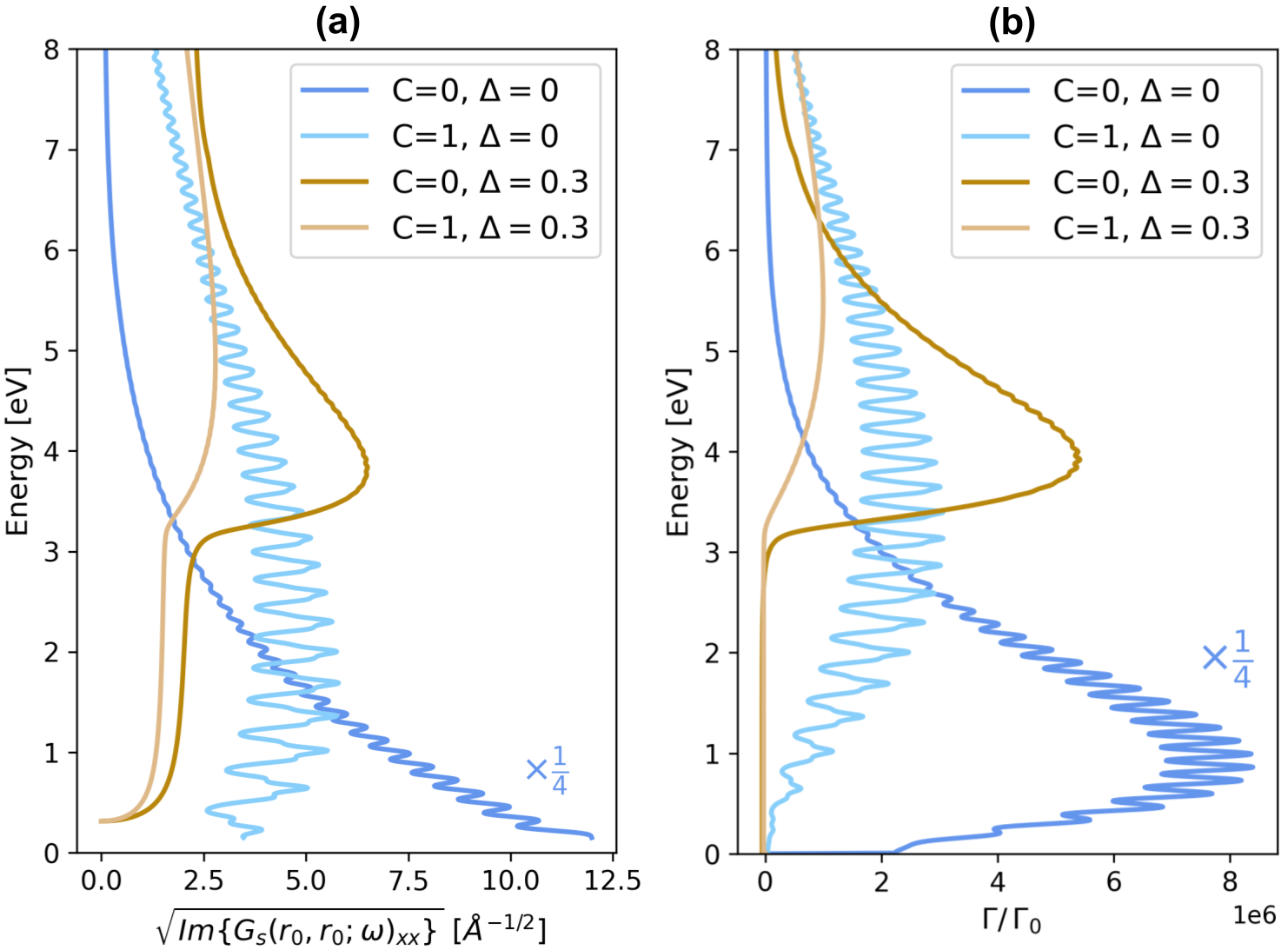}
    \caption{(a) Square root of $\mathrm{Im}G_s$ spectra for linear and dimerized chains of $N=500$ atoms, with and without Coulomb interactions. The emitter is placed above the chain at one-quarter of its length to probe a large number of bright resonances together. (b) Corresponding spontaneous emission rate enhancement (Purcell factor), $\Gamma/\Gamma_0$ for the same systems. Note that the response for the ($C=0,\,\Delta=0$) case is scaled by a factor of $1/4$. Near zero energy, the calculation of  $\mathrm{Im}G_s$ (Eq.~\ref{eq:Gs}) involves division of the simulated imaginary part of the scattered electric field by $\omega^2$. Consequently, any numerical noise in the electric field can become artificially amplified at low energies. To eliminate this artifact, the spectra below the first single-particle resonance are excluded.}
    \label{figs:long_chain}
\end{figure}
\begin{figure*}[!th]
\centering
\includegraphics[width=0.9\textwidth]{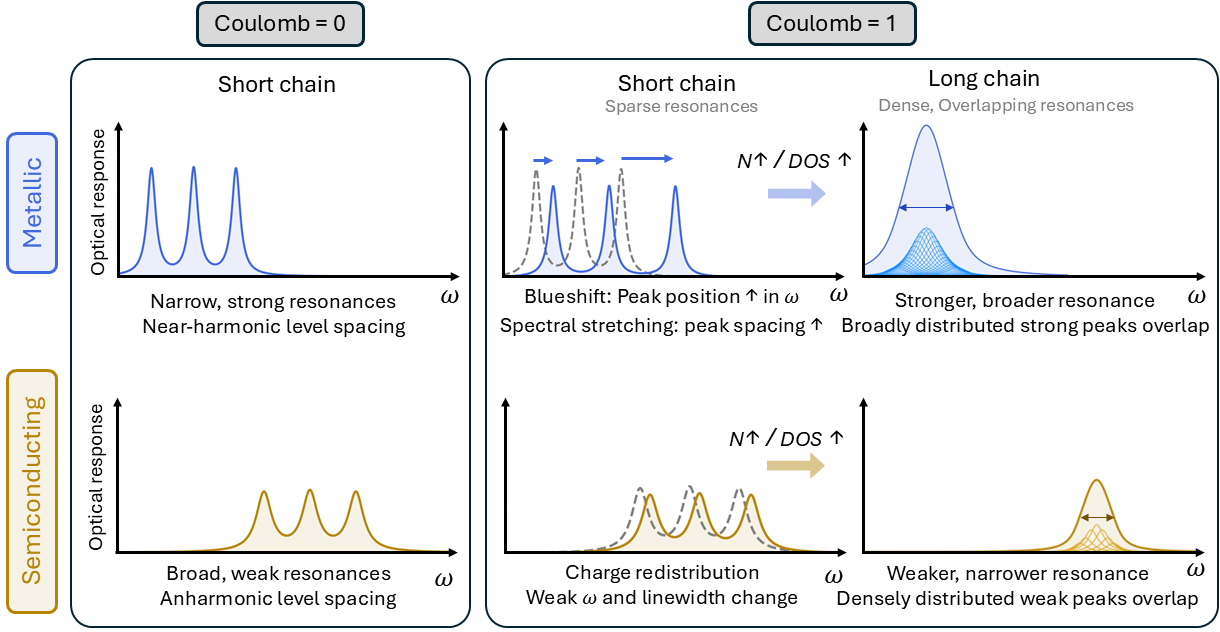}
    \caption{Scheme summarizing the impact of Coulomb interactions on metallic and semiconducting chains.}
\label{figs:coulomb_impact_scheme}
\end{figure*}
Its pronounced fast sloshing dynamics and the participation of multiple eigenstates are characteristic of a predominantly collective response, consistent with its high energy-based plasmonicity index, $\mathrm{EPI}=0.95$. 
The connection between these dynamical signatures and the resonance mechanism is discussed in Refs.~\cite{mueller2021,bryant2016}.
This collective mechanism of the resonance is characteristic of metallic structures. 
On the contrary, the dynamics shown in Fig.~\ref{figs:short_chain} (d) and (e) corresponds to semiconducting chain resonances [$\beta$ and $\gamma$ in Fig.~\ref{figs:short_chain}(b)]. Their smaller EPI values, $0.20$ and $0.10$, respectively, indicate a substantially weaker collective character. This can be seen from the slow evolution dominated by the Rabi oscillation of the charge between a pair of eigenstates, especially in panel (e). Both sloshing dynamics and Rabi cycles coexist in panel (d).

While short chains show features of both types of resonances, for longer ones, going towards the classical limit, the flat lower branch disappears, making the optical response predominantly collective for both types of chains, in agreement with results of our earlier study \cite{mueller2020}.   
The metallic chain retains a stronger response, and the larger Coulomb-induced spectral stretching produces a broader resonant band than in the semiconducting chain.

Figure~\ref{figs:long_chain} compares long metallic and semiconducting chains with $N=500$ atoms, calculated with and without Coulomb interactions. Panel (a) displays the square root of the relevant 
$\mathrm{ImG}_s$ spectra and captures the optical response of the chain as a scatterer, whereas the resulting Purcell factor shown in (b) accounts for the relaxation dynamics of a quantum emitter positioned in the vicinity of that scatterer. Mathematically, these quantities are related by a factor of $1/\omega$ (see Eq.~\ref{eq:tot_gamma}). 

Increasing the chain length reduces the spacing between electronic levels and produces a denser set of optical resonances. Their overlap at nonzero decoherence forms a broad spectral envelope. In semiconducting chains, the band gap shifts interband transitions to higher energies, while the anharmonic level spacing distributes their contributions over different frequencies.

Coulomb interactions reduce the resonance strength in both chain types, but aside from that their effects differ (Fig.~\ref{figs:long_chain}). In the metallic chain, the substantial blueshift and increased separation of resonance frequencies spread the strong resonances over a wider spectral range. At large $N$, their overlap produces a strong, broad resonant envelope (see the sketch in Fig.~\ref{figs:coulomb_impact_scheme}). In the semiconducting chain, Coulomb interactions primarily redistribute the induced charge and modify the spatial response, with much weaker spectral stretching. The overlapping resonances therefore produce a weaker and narrower envelope. The decoherence time used here was deliberately chosen to retain residual individual resonance peaks within the metallic envelope; their visibility reflects the finite linewidth relative to the resonance spacing.

\subsection{Beyond one-dimentional atomic chains}\label{real_chain}
\begin{figure*}[tbh]
    \centering
    \includegraphics[width=0.95\textwidth]{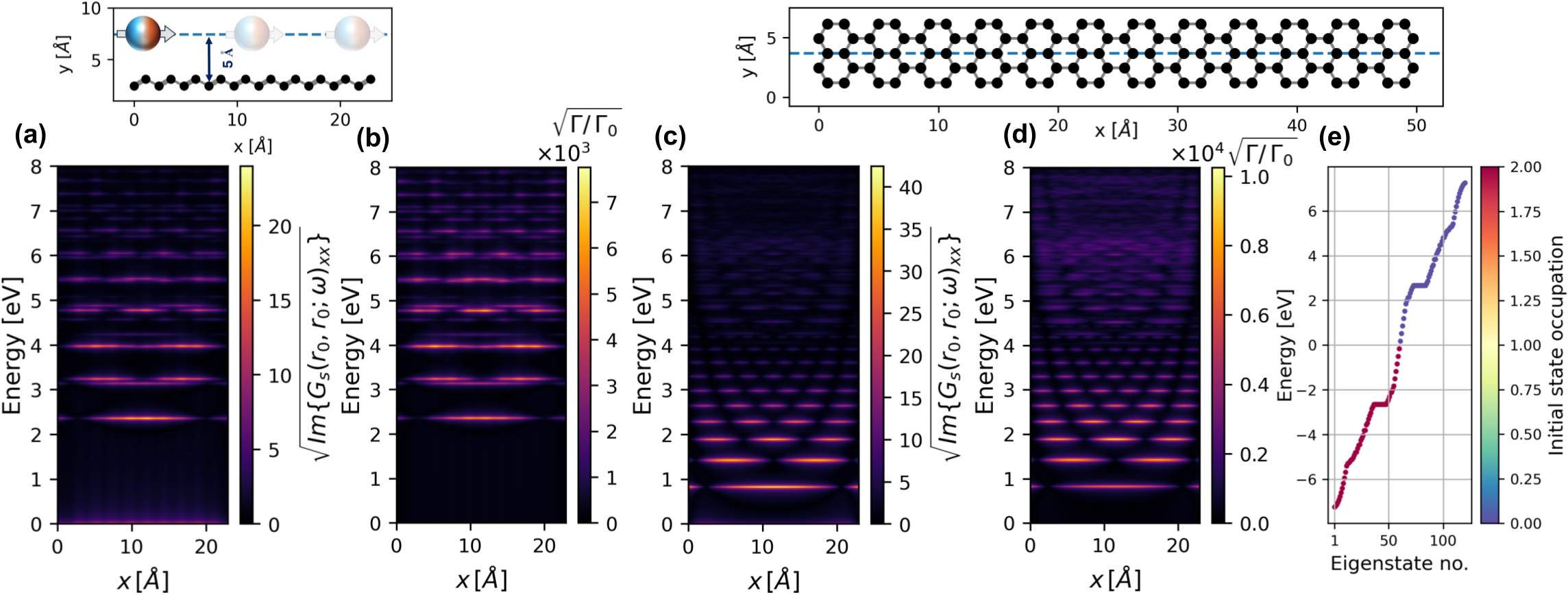}
    \caption{(a) Frequency-resolved imaginary part of the $x$-component of the scattered Green's function (square root), shown as a function of emitter position along the $x$-direction for a semiconducting trans-polyacetylene chain. The chain–emitter separation is fixed at $y = 5$ \AA, measured from the lower row of carbon atoms. (b) Frequency-resolved Purcell enhancement factor (square root) for the trans-polyacetylene system. (c) The same quantity as in panel (a) for a $5$ nm long metallic armchair graphene nanoribbon (5AGNR).
    The nanoribbon lies in the $xy$-plane, and the quantum emitter is scanned along the central longitudinal line (parallel to its length), maintaining a constant emitter–ribbon separation of $z = 5~\text{\AA}$ above the plane. (d) Frequency-resolved Purcell enhancement factor for the graphene nanoribbon system. (e) Energy spectrum of the graphene nanoribbon. The approximately harmonic states around the Fermi energy give rise to linear-chain-like spectral features. In contrast, the anharmonicity of states located further away from the Fermi energy produces dimer-chain-like characteristics, as reflected in the low- and high-energy regions of panels (c) and (d). For both systems, we set $C=1$.  }
    \label{figs:poly_gnr}
\end{figure*}
So far, our analysis has focused on systems that allow us to identify and study the fundamental mechanisms governing the engineering and modification of spontaneous emission using one-dimensional nanostructures. Equipped with this understanding of the underlying mechanisms, we now turn to more realistic systems and interpret the complex spectra of a trans-polyacetylene chain and an armchair graphene nanoribbon. Figure~\ref{figs:poly_gnr} shows the impact of these structures on the spontaneous emission of dipolar emitters placed at a distance of $5$~\AA\ from these structures. We plot the imaginary part of the Green's function, $\mathrm{Im}G_s$, in Fig.~\ref{figs:poly_gnr}(a), (c) for direct comparison with the calculations for the model systems described in the previous sections. Additionally, we present the spontaneous emission enhancement $\Gamma/\Gamma_0$ that can be observed in experiments [Fig.~\ref{figs:poly_gnr}(b), (d)]. The spectral features obtained for these systems closely resemble those found above for the chain models. This agreement suggests that the physical mechanisms identified in these models remain robust in realistic nanostructures. 

For modeling trans-polyacetylene we take the C--C $\pi$-bond hopping as $t_0=-2.66\,\mathrm{eV}$, and choose the intra- and intercell hoppings as $t_{\mathrm{intra}}=-2.8\,\mathrm{eV}$ and $t_{\mathrm{inter}}=-2.1\,\mathrm{eV}$, respectively, giving a single-particle band gap $(E_g=1.4\,\mathrm{eV})$, consistent with the experimentally reported 1.4--1.5 eV \cite{Fincher1979,Leising1988}. The chain geometry incorporates alternating C--C bond lengths of approximately 1.36~\AA\ and 1.44~\AA\ and a C--C--C bond angle of approximately $121^\circ $~\cite{baughman1985, Yannoni1983}.
The optical response retains the characteristic overlapping pattern known from the chain analysis. It additionally displays finer features associated with its zigzag geometry, which corresponds to different emitter-chain separations. The graphene nanoribbon, in turn, exhibits features characteristic of both the metallic and semiconducting chains \cite{Brey2006,Zheng2007}. Its energy spectrum [Fig.~\ref{figs:poly_gnr}(e)] resembles a mixture of the metallic- and semiconducting-chain patterns: nearly equally spaced energy levels in the energy range roughly $-2$ eV to $2$ eV, like the metallic chain, and immediately after that energy range, anharmonic energy level spacing like the semiconducting chain. The effect of this harmonic and anharmonic energy level spacing is directly reflected in its spontaneous emission properties: we observe the linear-chain-like enhancement profile of low-energy transitions [Fig.~\ref{figs:poly_gnr}(c), (d)] and the semiconducting-chain-like features at higher energy, i.e., a repeated metallic-chain-like pattern.
Carbon nanotubes provide a more complex example, with several spectral regions of increased density of electronic states related to van Hove singularities \cite{saito1998}. Their optical response and their influence on emission properties of quantum emitters can be explained by building upon the mechanisms described here for the basic models of metallic and semiconducting chains. 

The study above demonstrates that properties of more complex systems, even beyond linear chains, can be understood as combinations of characteristics originating in prototypical SSH chains. 
The optical response of these complex systems shows mixtures of features characteristic of both metallic and semiconducting chains, both regarding the spectra and spatial charge distributions. 

\section{Conclusions}
In this work, we have characterized the spontaneous emission enhancement of a quantum emitter in the vicinity of quasi-one-dimensional nanostructures, using the linear atomic chains as a prototypical model to uncover the underlying mechanisms. We have shown that the spontaneous-emission enhancement spectrum is resonant in nature: efficient enhancement occurs when the emitter is spectrally matched to the transition energies of the chain. By mapping the induced charge and scattered-field distributions, we also clarified why the emitter position plays a crucial role: Strong enhancement arises when the emitter couples efficiently to the induced charge dipoles of the chain.  A complementary perspective is provided by the induced-field patterns generated by these charge distributions.  Their detailed analysis reveals that spontaneous emission is quenched when the emitter is placed on nodal surfaces of the induced field.

The character of the spontaneous-emission enhancement is governed by the nature of the chain, whether metallic or semiconducting. 
 In metallic SSH chains, nearly uniform level spacing allows transition contributions to overlap and form strong resonances. 
This behavior is in line with the intuition from noble-metal plasmonics, where the concentration of many electronic transitions gives rise to intense collective resonances. 
In semiconducting chains, by contrast,
dimerization opens a band gap and introduces anharmonic spacing, shifting transitions to higher energies and distributing their contributions into weaker, broader resonance groups.

The spatial profile of the incident field provides an additional degree of control. While homogeneous illumination probes only dipole-allowed transitions, the inhomogeneous near field of the emitter breaks the mirror symmetry and couples to higher-order transition moments, keeping even-order and dimerization-suppressed modes optically active. The relative strength of these modes is set by wave-vector matching between the emitter field and the chain modes and by the mode-dependent decay of the scattered field, both tunable via the emitter–chain distance, chain length, and dipole orientation.

Coulomb interactions suppress resonance strength in both chain types but affect them differently. In short metallic chains, pronounced, mode-dependent blueshifts stretch the spectrum. In short semiconducting chains, interactions primarily separate nearly degenerate resonances into branches and redistribute induced charge, modifying their spatial profiles.
As the chains become longer, denser spectra promote overlap and collective responses in both cases. The stronger metallic resonances and greater Coulomb-induced spectral stretching nevertheless produce stronger and broader spontaneous-emission enhancement.

In realistic systems, the spontaneous emission generally combines the characteristics of both metallic and semiconducting chains. The conclusions drawn here therefore go beyond the SSH model and are directly relevant to realistic structures such as graphene nanoribbons, trans-polyacetylene chains, and carbon nanotubes. 
They also point to possible connections with local-probe experiments. The spatially resolved near-field response analyzed in this work may be relevant for the interpretation of scanning near-field optical microscopy measurements, where a nanoscale tip locally probes the optical response of a structure. As a further perspective, extending the present framework to include tip–sample tunnelling could establish links to STM-based probes, enabling electronic transport and optical near-field effects to be addressed within a common atomic-scale description.

\subsection*{Methods}
Numerical simulations were performed using the open software GRANAD \cite{dams2025}.
\subsection*{Author Contributions}

A.G. performed the numerical simulations, analytical calculations, and carried out the initial analysis of the results. D.D., A.G., and A.B. developed the numerical code. 
All authors contributed to the concept, discussions, analysis of results, and the final version of the manuscript.

\subsection*{Acknowledgments}
A.G., M.P. and K.S. acknowledge support from the National Science Centre, Poland (A.G. and K.S: Project No. 2020/39/I/ST3/00526; M.P.: Project No. 2024/55/D/ST3/02384). A.G., M.P and K.S. are grateful to the Donostia International Physics Center for their kind hospitality. 
D.D. and C.R. acknowledge support from the German Research Foundation under Grant No. RO 3640/14-1 within Project No. 465163297. 
A.A. and A.B. acknowledge funding from the Spanish Ministry of Science and Innovation (grants nos.PID2022-139230NB-I00, and TED2021-132074B-C32). A.A. also acknowledges the Gobierno Vasco UPV/EHU
(project no. IT-2032-26), the Departamento de Promoción Económica y Proyectos Estratégicos de la Diputación Foral de Gipuzkoa (Project AQUAS 2025-QUAN-000041-01), and the European Commission NaturSea-PV (GA 101084348). Research conducted in the scope of the Transnational Common Laboratory (LTC) Aquitaine-Euskadi Network in Green Concrete and Cement-based Materials. A. B acknowledges funding from project PID2025-170106NB-I00 funded by MICIU/AEI/10.13039/501100011033, and from Grant No. IT1904-26 from the Department of Science, Universities and Innovation of the Basque Government.” 

During the preparation of this manuscript, the authors used ChatGPT to assist with wording and improve the clarity and readability of the text. The authors reviewed and edited the AI-assisted text and take full responsibility for the final content of the manuscript.

\subsection*{Conflicts of Interest}

The authors declare no conflicts of interest.

\subsection*{Data Availability Statement}
The datasets generated and/or analyzed during the current study are available in the Research Data Repository of Nicolaus Copernicus University
repository at [PERSISTENT WEB LINK TO DATASETS].
\appendix
\section*{Appendices}
\counterwithin{figure}{section}
\section{Rotation of the quantum emitter}
\label{appendix:rotation}
\begin{figure}[tbh]
\centering
    \includegraphics[width=0.95\linewidth]{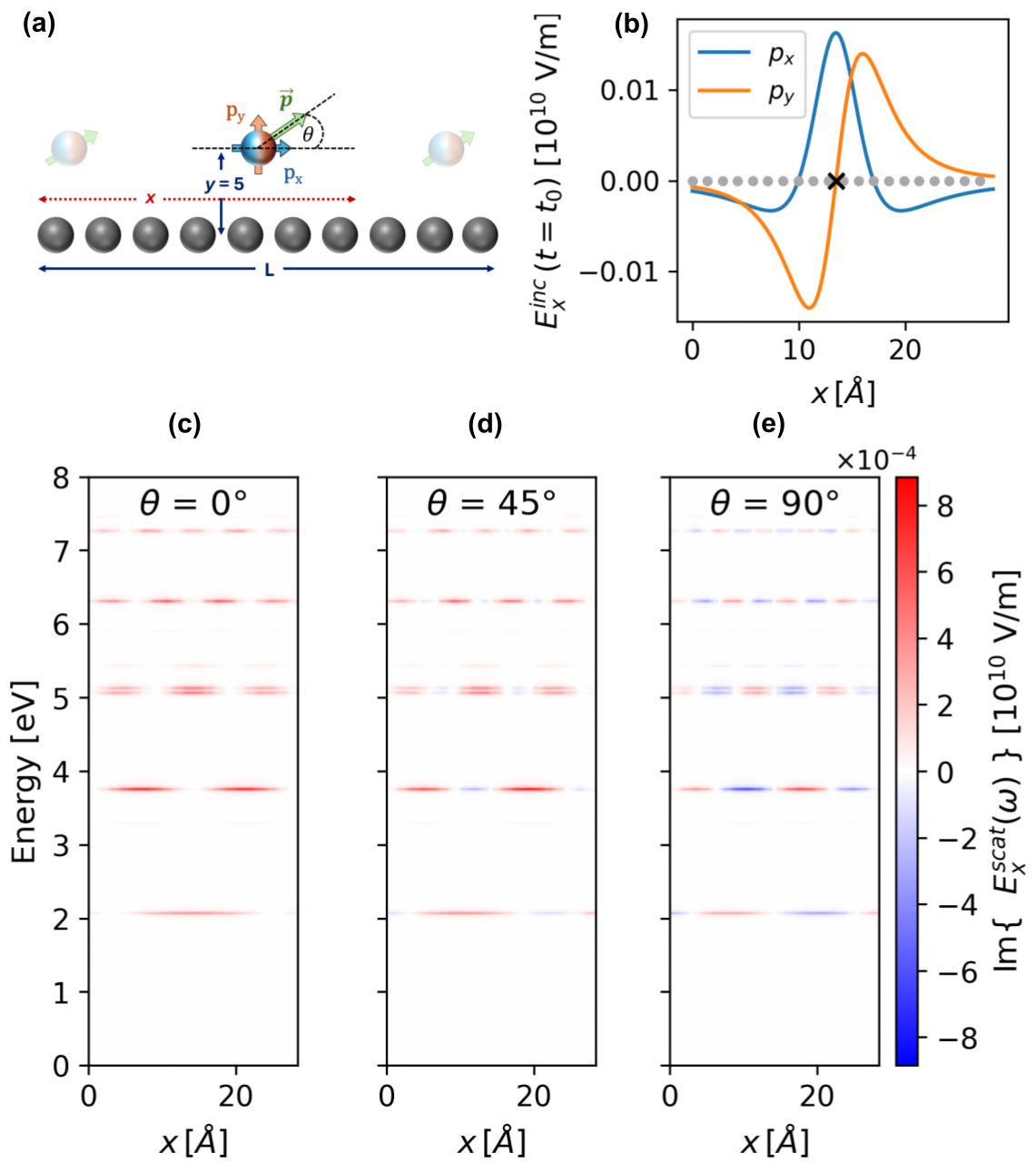}
    \caption{
    (a) Schematic illustration of the system configuration.
    (b) $x$-component of the incident electric field for a dipole located at the center of the chain, $5$~\AA\ from the chain axis (marked with a cross), with orientations parallel $(p_x)$ and perpendicular $(p_y)$ to the chain axis.
    Imaginary part of the $x$-component of the scattered field measured at the emitter position for dipole orientations rotated
    (c) $0^\circ$,  
    (d) $45^\circ$, and 
    (e) $90^\circ$
    relative to the $x$-axis.
    }
\label{figs:rotation}
\end{figure}

On the emitter side, the spontaneous-emission rate depends on both the position and orientation of the emitter transition dipole moment \cite{liu2014}. For various orientations $\theta$ (scheme in Fig.~\ref{figs:rotation}(a)), we analyze the imaginary part of the $x$-component of the electric scattered field $\mathrm{Im}\,E^\mathrm{scat}_{x}(\omega)$ [see Fig.~\ref{figs:rotation}(c)--(e)].
The scattered electric field is directly related to the scattered Green's tensor (Eq.~\ref{eq:Gs}) and provides more direct insight into the spatial symmetry of the modal structures with phase information.

The mirror symmetry of the chain yields identical responses for emitter positions at $x_0$ and $L-x_0$, producing spectra symmetric about the middle of the chain  [Fig.~\ref{figs:rotation}(c)].  
For a dipole oriented parallel to the chain ($\hat{p} \parallel \hat{x}$) and positioned sufficiently far from the chain edge, the incident field is approximately even about the point $x_0$ on the chain directly below the emitter, i.e., $E_x(x_0 - x) = E_x(x_0 + x)$ with $x\leq \frac{L}{2}-|x_0-\frac{L}{2}|$ [Fig.~\ref{figs:rotation}(a),(b)]. 
In contrast, for a perpendicular orientation ($\hat{p} \parallel \hat{y}$), the $x$-component of the incident field is approximately odd, and the corresponding scattered field changes sign, introducing additional nodal planes in the modal profiles. As shown in Fig.~\ref{figs:rotation}(e) ($\theta=90^\circ $), all the existing modes in Fig.~\ref{figs:rotation}(c) ($\theta=0^\circ $) are split into positive and negative parts separated by an additional nodal plane. This will be reflected in additional nodal lines in spontaneous emission enhancement plots. 
For other dipole orientations, the incident field can be expressed as a linear combination of even and odd components, and the resulting spectra reflect the relative contribution of each.  

Since the chain is finite, the incident field is not strictly even or odd, except for the emitter located at the chain center.  
However, the system retains approximate mirror symmetry, and
parity characteristics remain clearly observable in its response.

\section{Energy level splittings}
    \label{appendix: splitting}
\begin{figure}[t!]
    \centering
     \includegraphics[width=0.48\textwidth]{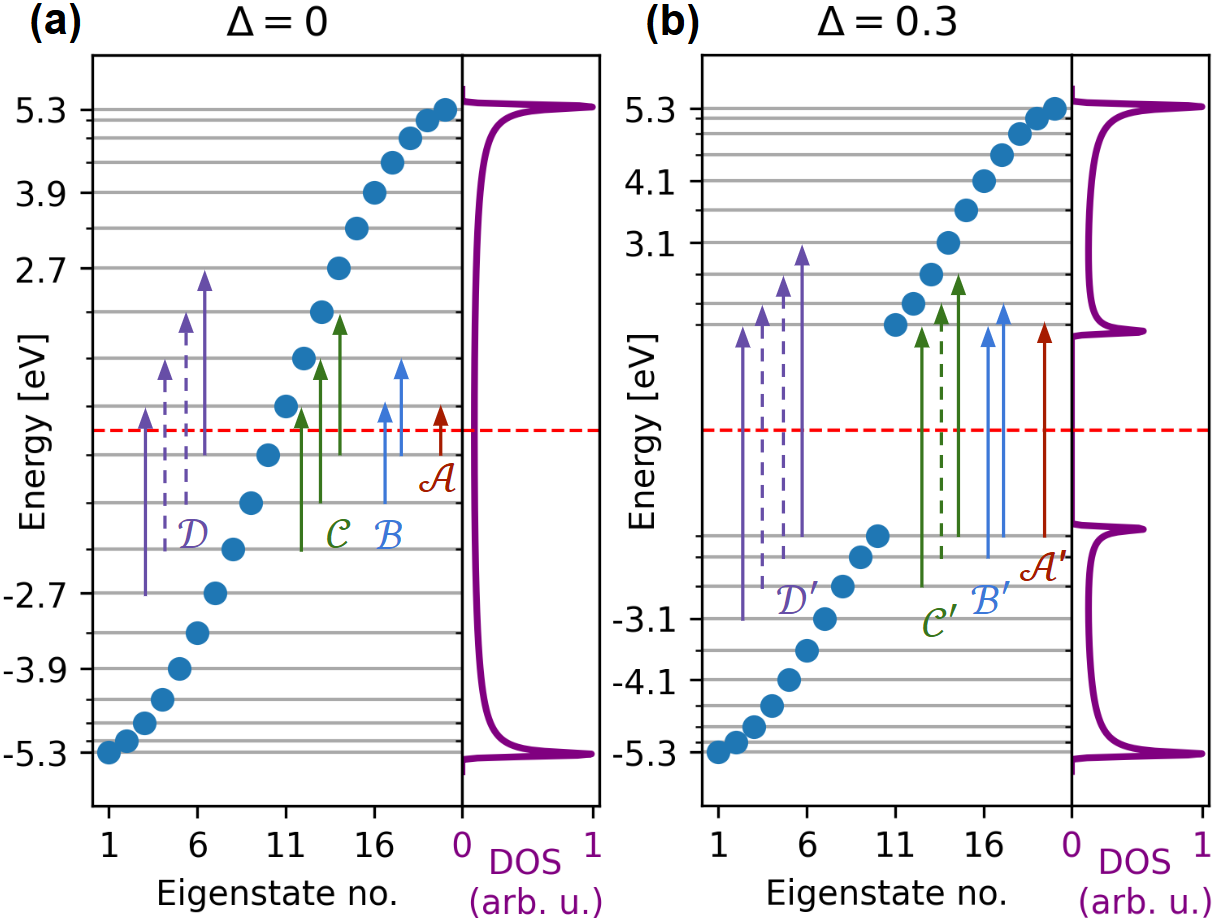}
    \caption{
    Energy level diagram of an $N=20$ atom
    (a) metallic chain ($\Delta = 0$) and (b) semiconducting chain ($\Delta = 0.3$).
    Each panel is accompanied by the density of states (DOS) of an infinite chain with the corresponding dimerization parameter.  
    Because the $N=20$ chain contains too few energy levels to represent the DOS reliably, the DOS of the infinite chain is shown instead, preserving the essential physics relevant to the discussion.
    The transitions associated with the first- to fourth-order modes in the non-interacting ($C=0$) spectra in Fig.~\ref{figs:delta-dep}(a,c) are indicated with arrows of different colours.
    }
    \label{figs:delta-dos}
\end{figure}
Dimerization of a linear chain opens a HOMO-LUMO energy gap and significantly modifies the spectral distribution of electronic states near the gap edges. This restructuring of the energy spectrum is reflected in the $ImG_s$ response as a splitting and redistribution of spectral modes. In the following, we relate these features to the underlying changes in the density of states induced by dimerization. Although the system is finite and thus characterized by discrete energy levels, the term “density of states” (DOS) is used in a generalized sense to describe the effective spacing and clustering of these levels in the energy space.

For sufficiently long metallic chains, the density of states near the Fermi level is approximately constant, which gives rise to a set of nearly degenerate low-energy transitions (transition set $\mathcal{C}$ in Fig.~\ref{figs:delta-dos}(a)) that appear as unresolved (effectively unsplit) modes in the $\mathrm{Im}G_s$ spectra, e.g., the third-order mode in Fig.~\ref{figs:delta-dep}(a). In contrast, higher-order transitions extend further from the Fermi level and probe deep into energy regions where the density of states varies. This varying density of states corresponds to non-equidistant energy levels referred to as  anharmonicity. 
As a result, transitions within a given order (e.g., transition set $\mathcal{D}$ in Fig.~\ref{figs:delta-dos}(a)) are non-degenerate and split spectrally, as observed for the fourth-order mode in Fig.~\ref{figs:delta-dep}(a).

Upon dimerization in semiconducting chains, the redistribution of states near the gap edges enhances the local energy-level density and introduces strong anharmonicity in the energy spectrum (DOS panel in Fig.~\ref{figs:delta-dos}(b)). As a result, previously unresolved transitions become spectrally distinguishable, leading to mode splitting.
Exceptions arise for the first- and second-order modes (transition sets $\mathcal{A}$, $\mathcal{B}$, and $\mathcal{A'}$, $\mathcal{B'}$ in Fig.~\ref{figs:delta-dos}).
The first-order mode involves only a single transition (HOMO $\to$ LUMO), therefore, cannot split. The second-order mode comprises two transitions that are symmetric about the Fermi level. As a result, the DOS modification introduced by dimerization affects both equally, and leaves them degenerate
and spectrally unsplit. 
In general, every even order mode $n$ consists of $\frac{n}{2}$ such symmetric pairs of transitions (e.g., dashed and solid marked transition pairs in the set $\mathcal{D}$ and $\mathcal{D'}$), whereas an odd-order mode $n$ contains $\frac{(n-1)}{2}$such pairs (e.g., solid marked transition pair in set $\mathcal{C'}$) along with an additional asymmetric transition (e.g., dash marked transition pair in set $\mathcal{C'}$), leading to $\lceil\frac{n}{2}\rceil$ branches for lower order modes.

For modes that are already split in the metallic chain, dimerization further modifies the underlying energy-level distribution and consequently alters the relative energies of the contributing transitions. This can lead to a reordering of the spectral branches associated with a given mode. In particular, the low-energy and high-energy branches of a given mode can interchange upon dimerization.
This behavior is illustrated in Fig.~\ref{figs:delta-dep}(a, c): analysis of the underlying transitions ($\mathcal{D}$ and $\mathcal{D'}$ in Fig.~\ref{figs:delta-dos}) shows that the transitions forming the upper branch of the fourth-order mode for $\Delta=0$ contribute to the lower branch for dimerization $\Delta=0.3$, indicating that the two spectral branches exchange their positions as $\Delta$ is varied continuously.

\section{Transition dipole moments evolution with dimerization}
\label{appendix:transition_moments}
\subsection{Finite chain}
\label{appendix:Finite_Chain}
\begin{figure}[tbh]
    \centering
    \includegraphics[width=0.5\textwidth]{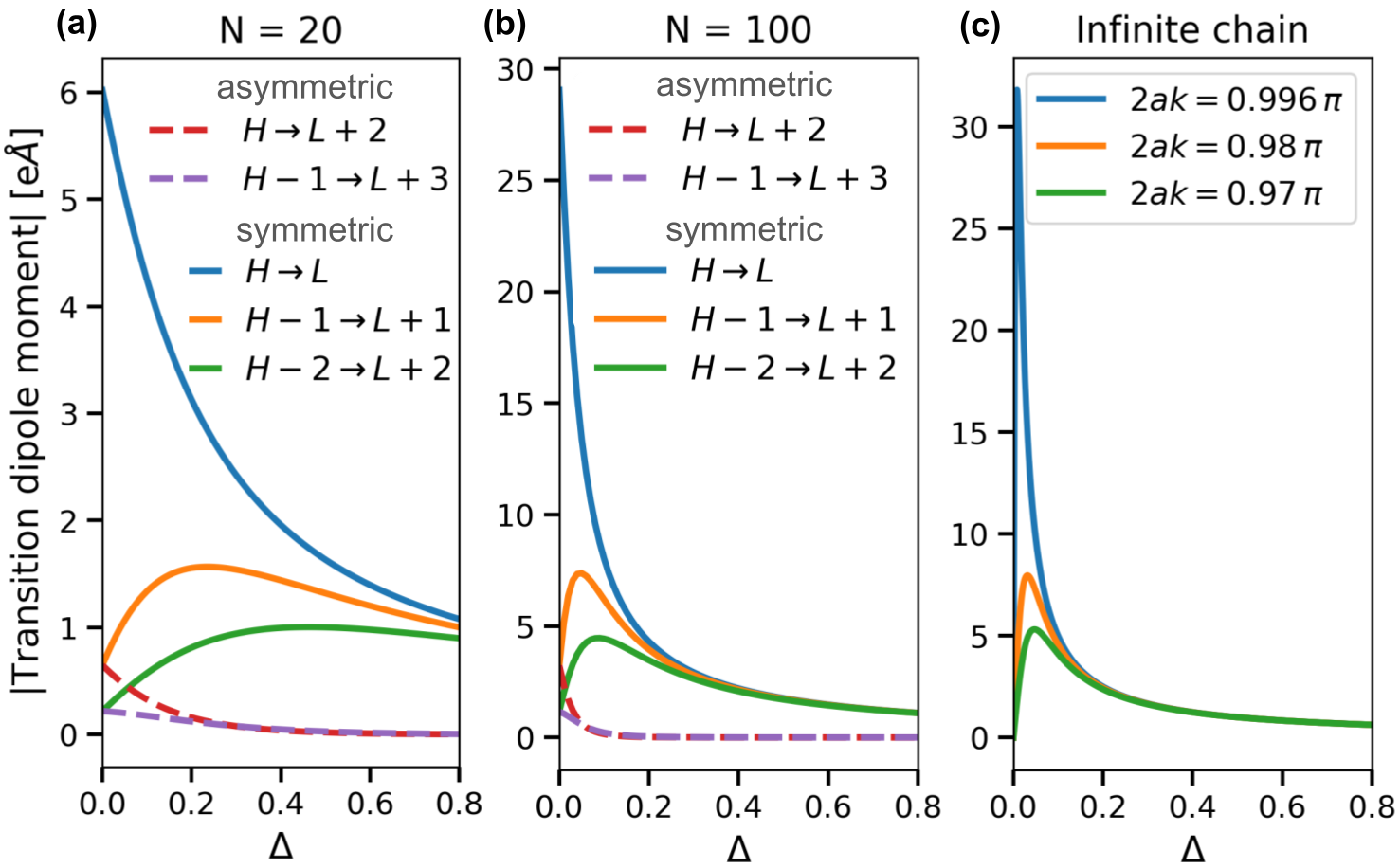}
    \caption{
    Selected transition dipole moments as a function of the dimerization parameter $(\Delta)$
    for a finite SSH chain with (a) $N=20$ atoms. (b) $N=100$ atoms. Solid lines correspond to transitions between states symmetrically positioned about the Fermi level, namely $(H\to L)$, $(H-1\to L+1)$, and $(H-2\to L+2)$, respectively, while the dashed lines denote the asymmetric transitions $(H\to L+2)$ and $(H-1\to L+3)$.
    (c)  Interband transition dipole moment of the infinite SSH chain evaluated at representative wave vectors$(k)$ near the Brillouin-zone boundary. As a transition takes place near the edge of the Brillouin zone $(k\to \frac{\pi}{2a}; k=0.996\frac{\pi}{2a} \text{ case})$, the transition dipole moment exhibits the same monotonic decay pattern as in the case of (HOMO $\to$ LUMO) transitions in a finite chain. The transitions deeper inside the Brillouin zone $(k=0.98\frac{\pi}{2a},\, k=0.97\frac{\pi}{2a})$ display the non-monotonic behavior characteristic of finite-chain symmetric transitions with increasing level spacing.
    }
    \label{figs:TDMs_num}
\end{figure}

To explain the mode quenching in absorption spectra with dimerization discussed in Sec.~\ref{subsec:disappearence_absorption_peak}, we examine the dipole moments of the relevant transitions as functions of the dimerization parameter $\Delta$ in Fig.~\ref{figs:TDMs_num}. We consider transitions between states symmetrically positioned about the Fermi level, with energy-level spacing between initial and final energy levels 1, 3, and 5 (e.g., $H\rightarrow L$, $H-1\rightarrow L+1$, and $H-2\rightarrow L+2$), as well as asymmetric transitions with energy-level spacings of 3 and 5 (e.g., $H\rightarrow L+2$ and $H-1\rightarrow L+3$). As the $\Delta$ value increases, the dipole moments of the asymmetric transitions are rapidly suppressed, whereas those of the symmetric transitions initially increase at small $\Delta$ and subsequently decrease, approaching finite values. The overall decrease is related to the state localization with a growing dimerization.
For these transitions, larger energy-level spacing corresponds to transitions farther from the Brillouin-zone boundary in the infinite-chain limit, and the peak transition dipole moment decreases accordingly.
The HOMO$\rightarrow$LUMO transition is an exception: its dipole moment is largest in the metallic limit ($\Delta\to0$) and decreases monotonically with increasing dimerization and saturates with other symmetric transitions.
A similar dependence of transition dipole moments on chain length and bond alternation has been obtained analytically for finite alternating polyene chains in Ref.~\cite{kohler1995}.

The behavior of the dipole moments for the symmetric transitions, in particular, the seemingly different HOMO-LUMO case, can be further understood by the analysis of the infinite chain limit.

\subsection{Infinite chain limit}
Here, we analyze infinite chains to identify which dipole moment characteristics originate from the universal
features of the SSH band structure, and which are related directly to the finite-size effects.

We consider an infinite SSH chain, with two sublattices, $s\in\{A, B\}$, per unit cell without spatial dimerization. The orbitals of the $n$-th unit cell are located at 
\begin{equation}
\label{eq:position_op}
  x_{n,s} = X_n+\xi_s, \quad X_n=2na,\quad \xi_A = -\frac{a}{2},\quad \xi_B = \frac{a}{2}  
\end{equation}
where $2a$ is the lattice constant with $a=1.42$~\AA\ interatomic distance. Within the tight-binding approximation, the position operator is $\hat{x} = \sum_{n,s} x_{n,s}\ket{n,s}\bra{n,s}$.

We use the Bloch basis
\begin{gather}
\label{eq:bloc_basis}
    \ket{k,s}=\sqrt{\frac{a}{\pi}}\sum_n e^{ikX_n} \ket{n,s}, \nonumber \\\mathrm{wave\,vector:\,}k\in \left(-\frac{\pi}{2a},\frac{\pi}{2a}\right].
\end{gather}

The Bloch phase specifies the unit-cell positions $X_n$ but does not specify the exact orbital positions within each cell, so both orbitals in a cell carry the same phase. An arbitrary state $\ket{\psi}=\int \mathrm{d}k \sum_s u_s(k)\ket{k,s}$ can be described in momentum space by the coefficient vector $\ket{u(k)}=(u_A,u_B)^T$, with $u_s(k)=\braket{k,s|\psi}$.   

The Bloch Hamiltonian in this $\ket{k,s}$ basis is
\begin{equation}
\label{eq:ssh_hamiltonian}
 H(k)=
\begin{pmatrix}
0  & h(k)\\
h^*(k)  & 0
\end{pmatrix}, \quad h(k)= t_1\,\,+\,t_2\,e^{-2ika}\quad ,
\end{equation}
where $t_1=t(1+\Delta)$ and $t_2=t(1-\Delta)$ are intra- and inter unit cell hopping rates respectively.

The eigenstates are
\begin{equation}
\label{eq:ssh_eigenstates}
\ket{u_\pm(k)}=\frac{1}{\sqrt{2}}
\begin{pmatrix}
\pm e^{i\phi(k)}\\
1
\end{pmatrix}, \quad \phi(k)=\arg \left (h(k)\right),
\end{equation}
where the $\pm$ sign corresponds to the energy of the state located, respectively, above or below the Fermi level.

In an infinite periodic system, the total dipole moment is not a well-defined quantity because of the spatially extended nature of Bloch states \cite{resta2007,ditler2023position}. Thus, optical transitions are characterized by momentum-resolved interband dipole matrix elements rather than by a global dipole moment \cite{blount1962,gu2013,margetis2023}. Additionally, the usual momentum space representation of the position operator $\hat x \to i\partial_k$ is not sufficient here as the two sublattices sit at different position within the unit cell carrying different phases which Bloch phase does not encode.

The components of $\hat x\ket{\psi}$ along the given Bloch basis are 

\begin{flalign}
\bra{k,s}\hat x\ket{\psi}=\sqrt{\frac{a}{\pi}}\sum_{n}(X_n+\xi_s)e^{-ikX_n}\braket{n,s|\psi}\nonumber\\
=(i\partial_k+\xi_s)u_s(k)
\end{flalign}

which gives action of $\hat{x}$ on $\ket{u(k)}$ as
\begin{flalign}
\hat{x}\to i\partial_k+\hat\xi, \qquad \hat\xi=\mathrm{diag}(\xi_A,\xi_B)=-\frac{a}{2}\sigma_z, \nonumber  \\
\sigma_z \text{: Pauli $z$ matrix}.
\end{flalign}

$\hat \xi$ is the embedding operator that specifies the phase of the two sublattices within a unit cell by indicating their exact positions. The exact form of $\hat\xi$ depends on the convention used in the definition of the Bloch basis.

The transition dipole moment amplitude between a pair of states in the valence and conduction
bands with momentum $k$ is
\begin{equation}
\begin{aligned}
\label{eq:correct d_cv }
|d_{cv}(k)|=e\left|\bra{u_+(k)}i\partial_k+\hat{\xi}\ket{u_-(k)}\right|
=e\left|\frac{1}{2}\partial_k\phi(k)+\frac{ a}{2}\right|\\
=\frac{ea}{2}\left| \frac{t_1^2-t_2^2}{t_1^2+t_2^2+2t_1t_2\,\cos(2ka)}\right|.
\end{aligned}
\end{equation}

All transitions involving states symmetrically positioned with respect to the Fermi level are analogous to direct interband transitions in the infinite SSH chain. The evolution of transition dipole moments with dimerization for different sets of transitions located near and far from the Brillouin zone boundary is shown in Fig.~\ref{figs:TDMs_num}(c). In contrast, transitions involving states asymmetrically located about the Fermi level in the finite chain correspond to indirect interband transitions in the infinite case, which are forbidden due to momentum conservation, as light carries a negligible amount of momentum. As shown in Fig.~\ref{figs:TDMs_num}, the dipole moments of these transitions decay rapidly with increasing chain length. In the thermodynamic limit, they vanish completely.

An important regime to examine is the metallic limit near the Brillouin zone boundary, with $\Delta\ll 1$ and near $k=\pm\frac{\pi}{2a}$, $\cos(2ka)\simeq-1+\frac{q^2}{2}$, with $q=(2ka\mp\pi),\, |q| \ll 1$.
\begin{equation}
    \label{eq:d_cv_metal}
    \left|d_{cv}(k\approx\pm\frac{\pi}{2a})\right|\simeq
    \frac{ea}{2}\left|\frac{4\Delta }{4\Delta^2+q^2}\right|
\end{equation}
describes the initial increase and the following decay of transition dipole moments. The dipole moment is maximized at $\Delta=\frac{|q|}{2}$, and then decays $\sim\frac{1}{\Delta}$.

In the limiting case, at the Brillouin-zone boundary, $q\to0$, the transitions are analogous to the HOMO $\to$ LUMO transitions in a finite chain. In this limit, the transition dipole moment $d_{cv} \sim \frac{1}{\Delta}$, reproducing the monotonic decay of the HOMO-LUMO transition dipole moment in the finite case.

Furthermore, the Lorentzian dependence on $q$ centered at the Brillouin-zone boundary $(q=0)$ with half-width $2\Delta$ implies that, for a particular chain (fixed $\Delta$), only transitions with $|q|\lesssim2\Delta$ carry a large dipole moment.

For $\Delta=0$, the chain becomes a uniform metal with lattice constant $a$. As we double the unit cell length to $2a$, the metallic band is folded into two branches, unlike two separate bands for $\Delta>0$ cases. Transitions between these folded bands are momentum-forbidden, resulting in $|d_{cv}|=0$. This does not imply that the metallic chain is optically inert. In the limit $\Delta\to0$, Eq.~\ref{eq:d_cv_metal} gives a singular distribution $|d_{cv}|=\frac{e\pi}{2}\delta(k-\frac{\pi}{2a})$, indicating the transition dipole concentrates at the Brillouin-zone boundary, where the folded branches become degenerate, and the transition frequency vanishes, $\omega\to0$.
This behavior is consistent with the classical Drude response of a metal. Drude conductivity for metals is given by $\sigma(\omega)=D\tau_{el}/(1-i\omega\tau_{el})$, with Drude weight $D$ and $\tau_{el}$ electronic relaxation time \cite{ashcroft1976}. In the ideal lossless limit, $\tau_{el}\to\infty$, its real part approaches $\mathrm{Re}\,\sigma(\omega)\to \pi D\delta(\omega)$, concentrating the optical response of an ideal metal at zero frequency.

\section{Spatial wave-vector matching and mode excitation}
\label{appendix:spatial_mode_matching}

 \begin{figure}[!tbh]
    \centering
        \includegraphics[width=\linewidth]{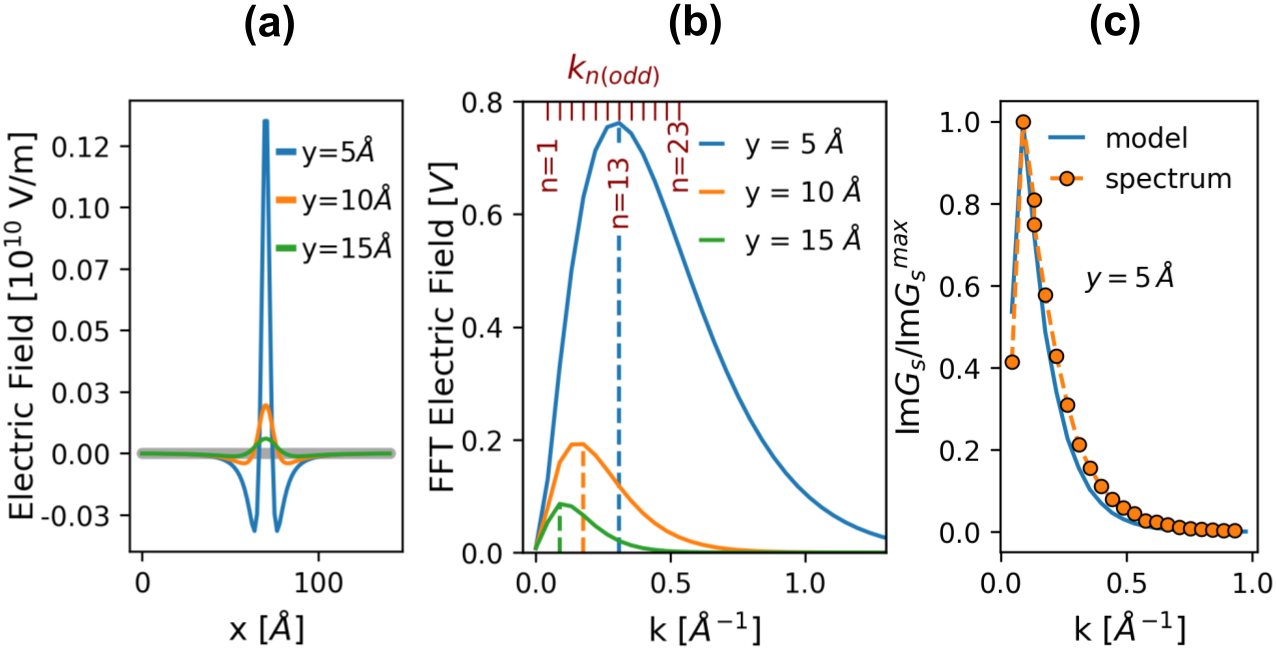}
        \caption{
        (a) $x$-component of the incident dipolar field along an $N=100$ atom metallic chain for emitter-chain separations of $5$~\AA\,, $10$~\AA\, and, $15$~\AA\, with the dipole positioned at the chain center.
        (b) Spatial Fourier transform of the incident field. The brown ticks on the upper axis indicate the wave vectors corresponding to the odd-order ($n$) modes supported by the metallic chain and observed in Fig.~\ref{figs:LC_xx}(b) for a dipole positioned at the chain center.
        (c) Comparison of the normalized $\mathrm{Im}G_s$ spectra obtained from the full spectral simulation and the wave-vector-matching model.
        The dots correspond to subsequent $k_{n(odd)}.$
        }
    \label{figs:incident_field_fft}
    \end{figure}

The inhomogeneous incident dipolar field on the chain $E_x(x)$ can be decomposed into spatial Fourier components $\tilde{E}_x(k) = \int_{-\infty}^{\infty} \mathrm{d}x\,E_x(x)e^{-ikx}.$
The resulting $k$-space distribution characterizes the wave vectors contained in the driving field. Figure~\ref{figs:incident_field_fft}(a) and (b) show the real-space field and its spatial Fourier spectrum for a source positioned above the middle of the chain and for several emitter-chain separations. As the emitter-chain separation decreases, the field becomes increasingly spatially inhomogeneous, and its spatial Fourier spectrum acquires larger-$k$ components.

The induced charge mode of a finite metallic chain can likewise be characterized by its spatial wave vectors. In the simplest approximation, the $n$-th mode can be represented by a sinusoidal spatial profile with characteristic wave vector $k_n=\frac{n\pi}{L}$, where $n$ is the mode index.  A mode is efficiently excited when the incident field has appreciable Fourier weight near its characteristic wave vector. 

For $\mathrm{Im}G_s$ spectra, the same emitter source acts as a detector of the scattered field generated by these induced charge modes. As the higher-order modes have greater charge inhomogeneity, their scattered field decays more rapidly with emitter-chain distance. Thus, in addition to excitation-side wave-vector matching, the detected response is subject to mode-dependent decay on the scattering side, which determines the final relative modal intensities detected in $\mathrm{Im}G_s$ spectra.
For instance, in the $N=100$ atom metallic chain, for the emitter at $y=5$~\AA\, the $13$th-order mode is excited most efficiently [Fig.~\ref{figs:incident_field_fft}(b)], but since the spatial decay rate of this mode is very high, in the final $\mathrm{Im}\,G_s$ spectra, the third-order mode with much lower decay has the maximum intensity [see Fig.~\ref{figs:incident_field_fft}(c) and intensity along dashed line in Fig.~\ref{figs:LC_xx}(b)].
Thus, a mode can be selectively and efficiently excited by varying either the chain length, which changes the supported $k_n$, or the emitter-chain separation, which modifies the $k$-space distribution of the incident field and the spatial decay of the corresponding scattered field.

\bibliographystyle{unsrtnat}
\bibliography{controling}

\end{document}